\documentclass{sig-alternate}
\usepackage{mathptmx} % use Times font

\usepackage{fancyhdr}
\usepackage[normalem]{ulem}
\usepackage[hyphens]{url}
\newcommand{\subparagraph}{} % IEEEtrans class doesn't define the subparagraph level
\usepackage{titlesec}
\usepackage[utf8]{inputenc}
\usepackage{booktabs}
\usepackage{tabulary}
\usepackage{amsmath}       
\usepackage{amsfonts}
\usepackage{nicefrac}
\usepackage{enumitem}
\usepackage{todonotes}
\usepackage{alltt}
\usepackage{algorithm}
\usepackage{xspace}
\usepackage{etoolbox}
\usepackage{graphicx}
\usepackage{array}
\usepackage{tabularx}
\usepackage{multirow}
\usepackage{adjustbox}
\usepackage{enumitem}
\usepackage{lipsum}
\usepackage{hyphenat}
\usepackage{listings}
\usepackage{color} 
\usepackage{breqn}
\usepackage{pifont}
\usepackage[sort,nocompress]{cite}
\usepackage[final]{microtype}
\usepackage{flushend}
\usepackage[bookmarks=true,breaklinks=true,letterpaper=true,colorlinks,linkcolor=black,citecolor=blue,urlcolor=black]{hyperref}

\newcommand{\ignore}[1]{}
\newcommand{\mech}{CODA}

\newcommand{\figwidth}{3.3in}

\title{CODA: Enabling Co-location of Computation and Data for Near-Data Processing}
\author{
  Hyojong Kim$^\dagger$\quad Ramyad Hadidi$^\dagger$\quad Lifeng Nai$^\dagger$\quad Hyesoon Kim$^\dagger$\quad \vspace{1ex} \\
  Nuwan Jayasena$^\star$\quad Yasuko Eckert$^\star$\quad Onur Kayiran$^\star$\quad Gabriel H. Loh$^\star$ \\
  \\
  Georgia Institute of Technology$^\dagger$\quad AMD Research$^\star$ \vspace{1ex}
}

\begin{document}
\maketitle
\pagestyle{plain}

\begin{abstract}

Recent studies have demonstrated that near-data processing (NDP) is an effective technique for improving performance and energy efficiency of data-intensive workloads. 
However, leveraging NDP in realistic systems with multiple memory modules introduces a new challenge. 
In today's systems, where no computation occurs in memory modules, the physical address space is interleaved at a fine granularity among all memory modules to help improve the utilization of processor-memory interfaces by distributing the memory traffic. 
However, this is at odds with efficient use of NDP, which requires careful placement of data in memory modules such that near-data computations and their exclusively used data can be localized in individual memory modules, while distributing shared data among memory modules to reduce hotspots. 
In order to address this new challenge, we propose a set of techniques that (1) enable collections of OS pages to either be fine-grain interleaved among memory modules (as is done today) or to be placed contiguously on individual memory modules (as is desirable for NDP private data), and (2) decide whether to localize or distribute each memory object based on its anticipated access pattern and steer computations to the memory where the data they access is located. 
Our evaluations across a wide range of workloads show that the proposed mechanism improves performance by 31\% and reduces 38\% remote data accesses over a baseline system that cannot exploit computate-data affinity characteristics.
\end{abstract}
\section{Introduction}
\label{sec:introduction}

Recent studies have demonstrated that near-data processing (NDP) is an effective technique to improve performance and energy efficiency of data-intensive workloads~\cite{ahn:yoo15,ahn:hon15,aki:fra15,zha:jay13,zha:jay14,chu:jay13,hsi:ebr16,nai:had17}. 
However, leveraging NDP in realistic systems with multiple memory modules (e.g., DIMMs or 3D-stacked memories) introduces a new challenge. 
In today's systems, where no computation occurs in memory modules, the physical address space is interleaved at a fine granularity among all memory modules to help improve the utilization of processor-memory interfaces by distributing the memory traffic. 
However, this is at odds with efficient use of NDP, which requires careful placement of data in memory modules such that near-data computations and the data they exclusively use can be localized in individual memory modules, while distributing shared data among memory modules to reduce memory bandwidth contention. 

\begin{figure*}
\centering
\includegraphics[width=\textwidth,keepaspectratio]{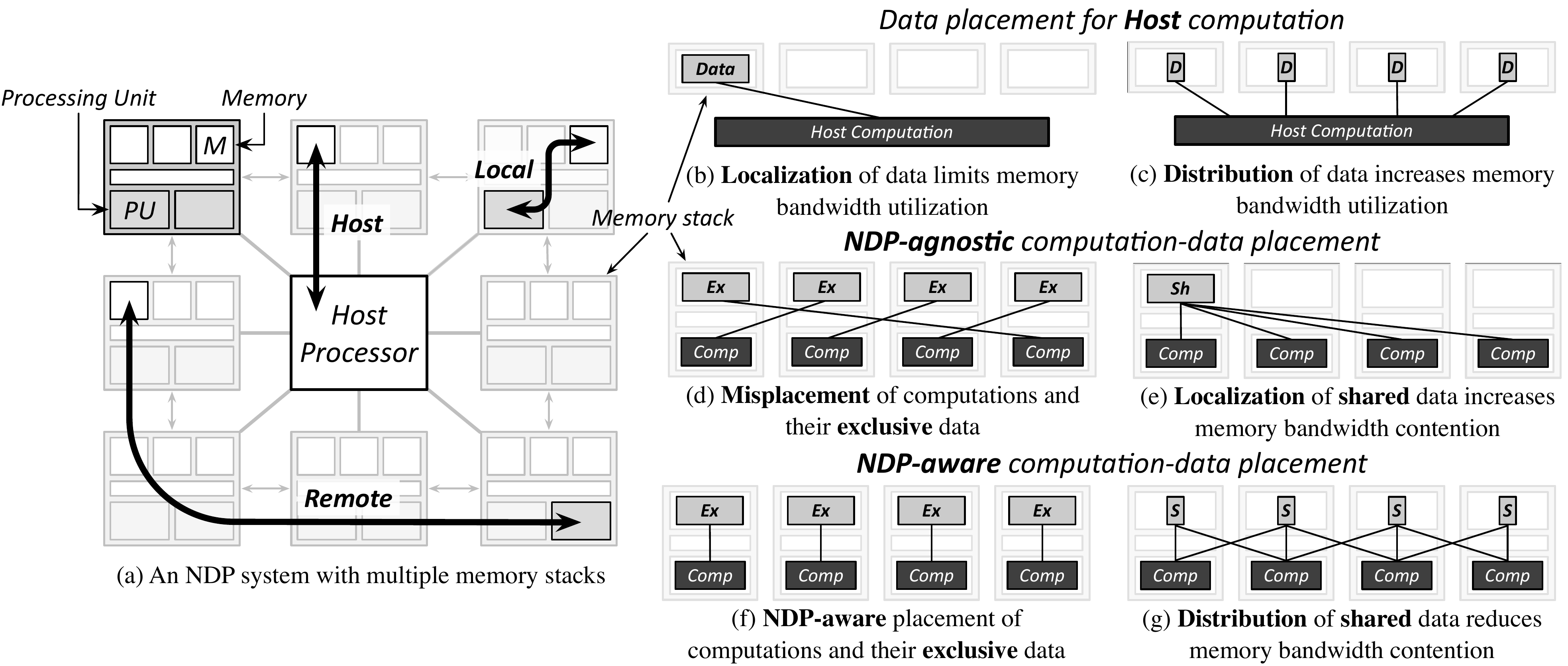}
\caption{(a) Overview of a near-data processing system with multiple memory stacks. (b) A localization of data in one memory stack that limits memory bandwidth utilization for host computation. (c) A distribution of data across memory stacks that increases memory bandwidth utilization for host computation. (d) An NDP-agnostic placement of computations and their exclusively used data that increases remote data accesses. (e) Localization of shared data in one memory stack that increases memory bandwidth contention. (f) An NDP-aware placement of computations and their exclusively used data that eliminates remote data accesses. (g) Distribution of shared data acorss memory stacks that reduces memory bandwidth contention.}
%\vspace{-1em}
\label{fig:baseline_architecture_motivation}
\end{figure*}

Figure~\ref{fig:baseline_architecture_motivation} (a) shows a high-level diagram of an NDP system with multiple memory stacks.
The system consists of the host processor and multiple 3D memory stacks, where each of which has one or more processing units on its logic layer.
Memory stacks are connected with the processor-centric topology proposed by Kim et al.~\cite{kim:kim13}, constituting the entire memory address space; they are used as main memory for all processing units in the system, including the host processor and processing units in memory stacks.
While processing units in memory stacks can transparently access data in any memory stack, accesses to data in other memory stacks use the low bandwidth off-chip links and traverse the interconnect, incurring higher latency and leading to lower system performance and energy efficiency.
On the other hand, a local data access, which occurs when a processing unit accesses data in its local memory stack, utilizes high memory bandwidth within the memory stack, incurring lower latency and leading to higher system performance and energy efficiency.
Therefore, it is critical to minimize remote data accesses for the efficient use of NDP. 

Figure~\ref{fig:baseline_architecture_motivation} (b) and (c) demonstrate the need for the distribution of data across memory stacks to increase the bandwidth utilization of processor-memory interfaces when computation is performed in the host processor.
Figure~\ref{fig:baseline_architecture_motivation} (d) - (g) demonstrate the need for more careful placement of computations and data when computation is performed in the processing units near memory.
Figure~\ref{fig:baseline_architecture_motivation} (d) and (e) present two cases of NDP-agnostic computations and data placements.
In Figure~\ref{fig:baseline_architecture_motivation} (d), computations and the data they exclusively use are placed in different memory stacks, causing all the remote memory traffics among memory stacks.
In Figure~\ref{fig:baseline_architecture_motivation} (e), shared data is localized in one memory stack, increasing memory bandwidth contention in that memory stack.
Figure~\ref{fig:baseline_architecture_motivation} (f) and (g) present the ideal cases of NDP-aware computations and data placements.
In Figure~\ref{fig:baseline_architecture_motivation} (f), computations and their private data are placed in the same memory stacks, eliminating all the remote data accesses and leading to better use of NDP.
In Figure~\ref{fig:baseline_architecture_motivation}(g), shared data is spread across memory stacks, reducing memory bandwidth contention in the first memory stack and leading to performance improvement.

Our \textbf{goal} in this paper is to realize such an NDP-aware placement of computation and data with very low overhead, while not sacrificing the host processor performance;
Near-data computations and the data they exclusively use should be localized in individual memory stacks for efficient use of NDP, whereas shared data and data accessed by the host processor should be spread across memory stacks to reduce memory bandwidth hotspots and to maximize bandwidth utilization.
Unfortunately, there are two key \textbf{challenges} that need to be solved to achieve this goal: (1) how to \textit{selectively} localize data in a system with multiple memory stacks where address space is finely interleaved (data is spread across multiple memory stacks by default), and (2) how to \textit{identify} the data that favors localization and how to \textit{co-locate} computations with the data they exclusively use?

(1) To solve the first challenge, we propose a lightweight hardware mechanism that supports dual-mode address mapping at a page granularity, so that a page can be spread across memory stacks or localized to a single memory stack.
The key idea is to use different sets of address mapping bits for each memory page depending on its anticipated access pattern, allowing the two sets of mappings to co-exist; low order bits are used to distribute a page across memory stacks, whereas high order bits are used to place (or localize) an entire page in a single memory stack.
The granularity information for each memory page is stored in the page table entry (PTE) and translation lookaside buffer (TLB) entry.
At the time a virtual address is translated into a physical address and the memory request is sent, our mechanism uses the appropriate address mapping depending on the granularity information.
Admittedly, the concept of changing address mapping to change data layout or to increase memory-level parallelism is not new~\cite{zha:zhu00,gha:jal16}.
However, our proposed mechanism is different from previous proposals in that it enables coexistence of pages with different address mappings while not requiring large-scale page migrations.

%\footnote{We use the term thread to refer to work-item in OpenCL and thread in CUDA.}
(2) Identifying exclusively accessed data and co-locating computations that use that data is particularly difficult for GPU systems for two reasons.
First, data structures are usually allocated (and often initialized) by the host processor (CPU) before kernel invocation and used by all threads in the kernel later.
Which thread accesses which (and which part of) data structures is not determined at the time data structures are allocated.
Second, and more importantly, thread-blocks\footnote{We use the term thread-block to refer to work-group in OpenCL and block in CUDA.} can be scheduled to any core in GPU systems. 
Considering the efficacy of an NDP system depends on the co-location of thread-blocks and the data they exclusively use; this \textit{nondeterministic} aspect of GPU execution models hinders the efficient use of NDP.
For these reasons, we target a GPU-based NDP system. 
The applicability of our mechanism to other core types is discussed later.

To solve the second challenge in such a GPU-based NDP system, we make two key observations.
First, the amount of data used by one thread-block is often determined by the number of threads in a thread-block and the amount of data each thread accesses. 
The latter can be estimated by either compile-time analyses (for input-independent access patterns) or profiler-assisted techniques (for input-dependent access patterns). 
Although the number of threads in a thread-block is often input-dependent, it is determined before kernel invocation (specifically, even before data structures are allocated).
Both combined, we come to a conclusion that the amount of data used by one thread-block \textit{can} be estimated.
Second, we observe that a slight restriction on the thread-block scheduling policy based on the affinity between thread-blocks and compute resources (memory stacks) in a heterogeneous system in terms of memory access latencies enables the efficient use of NDP despite the potential compute resource sub-utilization or load imbalance.
We base our observation on the fact that the number of threads and thread-blocks is typically much greater than the number of cores in GPU systems.
Based on these observations, we propose a software/hardware cooperative solution that (1) utilizes a compiler- and profiler-based technique to analyze the access pattern for each memory object and determine how each memory object should be layered across memory stacks, and (2) uses an affinity-based scheduling algorithm to steer thread-blocks to the memory stack where the data they access is located. 

Our paper makes the following \textbf{contributions}. 
First, we propose a lightweight hardware mechanism that supports dual-mode address mapping at a page granularity, such that a page can be spread across memory stacks or localized to a single memory stack.
This enables pages with different address mappings to coexist in the same memory space.
Second, we propose a software/hardware cooperative solution that utilizes a compiler-based and profiler-assisted technique to decide whether to localize or distribute each memory object based on its anticipated access pattern.
This mechanism steers computations to the memory where data they exclusively access is located, thereby achieving efficient use of NDP.
Third, we evaluate our proposed mechanism with a wide range of data-intensive workloads and show that it improves performance by 31\% and reduces 38\% of remote data accesses over a baseline system that does not have dual-mode address mapping nor an affinity-based computation and data co-placement mechanism.
\section{Background}
\label{sec:background}

\subsection{Baseline Architecture}
\label{subsec:baseline}

%\vspace{-0.75em}
\begin{figure}[htb]
\includegraphics[width=\columnwidth,keepaspectratio]{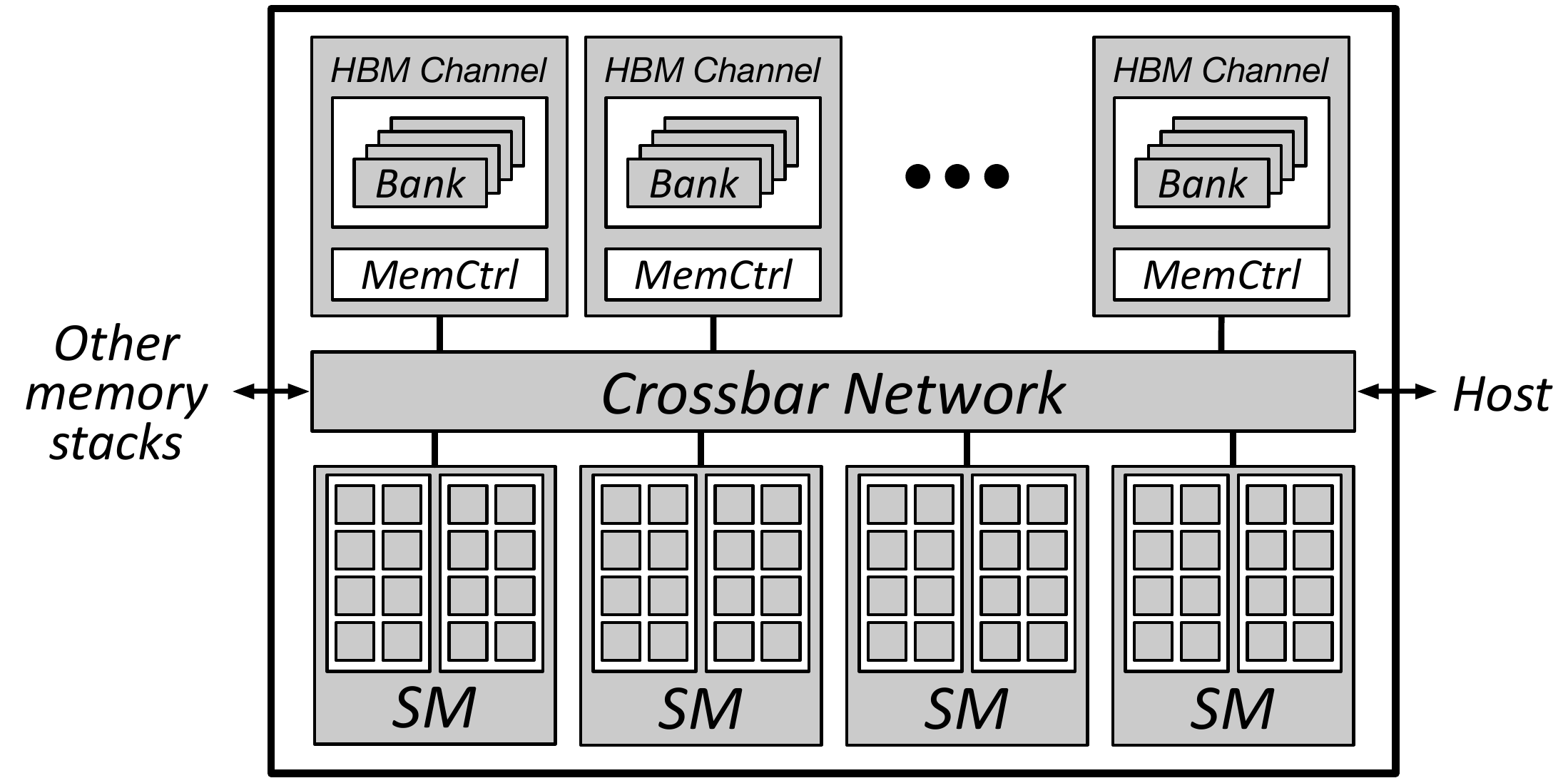}
\caption{Overview of an NDP memory stack}
\label{fig:baseline_memory_stack}
\end{figure}
%\vspace{-0.5em}

Figure~\ref{fig:baseline_memory_stack} shows a conceptual diagram of an example GPU-based NDP memory stack.
Although our mechanism does not rely on any particular memory organization, we choose high bandwidth memory (HBM) as our baseline~\cite{hbm2}.
HBM is composed of multiple memory channels and uses a wide-lane bus interface to achieve high memory bandwidth and low power dissipation. 
Each memory stack has one or more streaming multiprocessors (SMs) on its logic layer and high-speed off-chip links for remote data accesses - to/from other memory stacks and the host processor - and a crossbar network that connects SMs and HBM.
We assume SMs in the memory stack are equipped with a hardware TLB and memory management units (MMUs) that access page tables and are capable of performing virtual address translation.

\subsection{Programming Model}
We use the widely adopted GPU programming model as our programming model. 
The host processor launches GPU kernels, and the runtime system partitions and distributes thread-blocks across all the SMs in the system.
Each HBM has multiple (four in our evaluation) SMs, so up to the number of SMs $\times$ the number of thread-blocks per SM are concurrently executed in each memory stack. 

\subsection{Networks}
\label{subsec:network}
There are three kinds of networks in our system: (1) a network among the host processor and memory stacks (denoted as \textbf{Host} in Figure~\ref{fig:baseline_architecture_motivation}), (2) a network among memory stacks (denoted as \textbf{Remote} in Figure~\ref{fig:baseline_architecture_motivation}), and (3) a network that connects SMs in a memory stack to their local memory (denoted as \textbf{Local} in Figure~\ref{fig:baseline_architecture_motivation}). 
To provide an efficient execution for legacy (non-NDP-aware) applications, it is logical to dedicate most of the network resources available in a system for the host processor and memory stack connections. 
For this reason, we assume that the Remote network has much less bandwidth than the Host network. 
The order of memory bandwidth among the three types of networks is as follows: Local > Host > Remote.  

\subsection{Address Interleaving}

To increase memory-level parallelism, or to reduce channel/rank/bank conflicts, fine-grain interleaving is typically used in modern memory systems by striping small chunks of the physical address space (often the size of a few cache lines) across different banks, ranks and channels. 
In a system with multiple memory stacks, a page can be striped across multiple stacks with fine-grain interleaving, or the entire page can be allocated in a single memory stack with coarse-grain interleaving.
Complex address decoding schemes have been studied before~\cite{zha:zhu00,pet:dan16}; however, for brevity, we assume a simple address mapping scheme.
We discuss the applicability of our mechanism in systems with complex address mapping schemes in Section~\ref{subsec:complex_address_mapping}.

%\vspace{-0.5em}
\section{Motivation}
\label{sec:motivation}

%\vspace{-0.75em}
\begin{figure}[htb]
\includegraphics[width=\columnwidth,keepaspectratio]{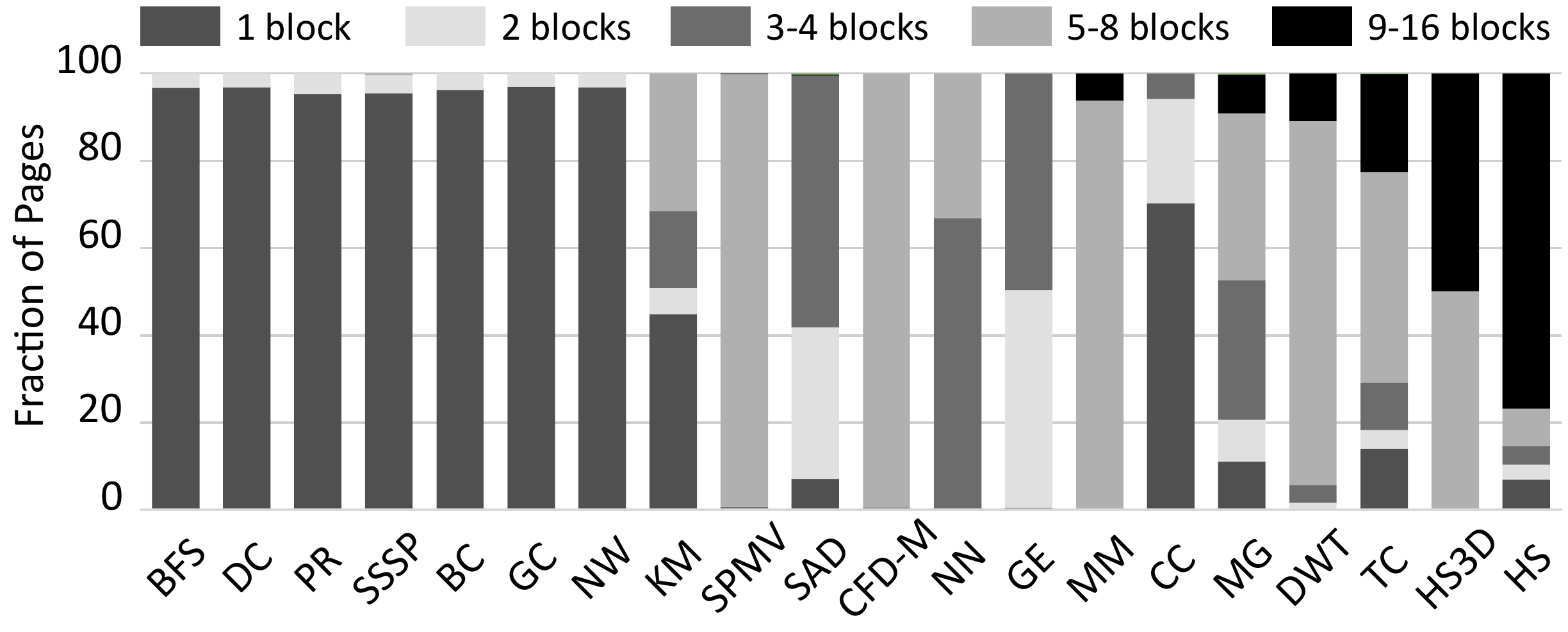}
\caption{Distribution of memory pages according to the number of thread-blocks that access each page}
\label{fig:intro_motivation}
\end{figure}
%\vspace{-0.5em}

Figure~\ref{fig:intro_motivation} shows distribution of memory pages according to the number of thread-blocks that access each memory page for various data-intensive workloads from publicly available GPU benchmark suites~\cite{lif:yin15,che:boy09,parboil}.
It is observed that for some workloads, such as \texttt{BFS}, \texttt{DC}, \texttt{PR}, \texttt{SSSP}, \texttt{BC}, \texttt{GC}, and \texttt{NW}, most pages are accessed by only one or two thread-blocks. 
In traditional systems, where no computation occurs in memory, distributing pages irrespective of which and how many thread-blocks access them helps improve the utilization of processor-memory interfaces by distributing the memory traffic.
However, when computation is performed \textit{near} memory (as is enabled by NDP), distributing such pages across memory stacks incurs lots of remote traffics.
Therefore it is imperative to place such pages (exclusively used data) and the thread-blocks (computations) that access them in individual memory stacks for efficient use of NDP.
In contrast, in the case of \texttt{HS3D} and \texttt{HS}, most pages are accessed by almost all thread-blocks.
Even in the presence of NDP, it is better to distribute such pages (shared data) across memory stacks to reduce memory bandwidth contention.

From this, we make two observations.
First, some pages are accessed exclusively by a few thread-blocks, while other pages are accessed, or shared, by many thread-blocks.
The exclusively used pages should be placed in individual memory stacks with the thread-blocks that access them to eliminate remote data accesses, and the shared pages should be distributed across memory stacks to reduce memory bandwidth contention.
Second, each application has different distribution of exclusive and shared pages.
For example, most pages in \texttt{BFS} are exclusively used, so the memory system should be capable of localizing all of them.
On the other hand, most pages in \texttt{HS} are shared, so the memory system should also be capable of distributing all of them.
These observations motivate the need for a mechanism that can allocate localized pages versus distributed pages \textit{flexibly} based on an application's needs.
\section{Mechanism}
\label{sec:mechanism}

In this section, we describe our mechanisms to enable co-location of computations and data in a system with multiple NDP memory stacks. 
Section~\ref{subsec:mechanism:overview} provides an overview of a non-NDP-aware distribution of computations and data, and demonstrates how an NDP-aware mechanism can improve it.
Section~\ref{subsec:address_interleaving} describes a hardware mechanism that supports dual-mode address mapping at a page granularity that either distributes a page across memory stacks or localizes the page to a single memory stack.
Section~\ref{subsec:algorithm} describes a software/hardware cooperative solution that utilizes a compiler-based and profiler-assisted technique to decide whether to localize or distribute each memory object based on its anticipated access pattern and an affinity-based scheduling algorithm that realizes steering of thread-blocks to the memory stack where the data they access is located. 

\subsection{Overview}
\label{subsec:mechanism:overview}

\begin{figure}[htb]
\includegraphics[width=\columnwidth,keepaspectratio]{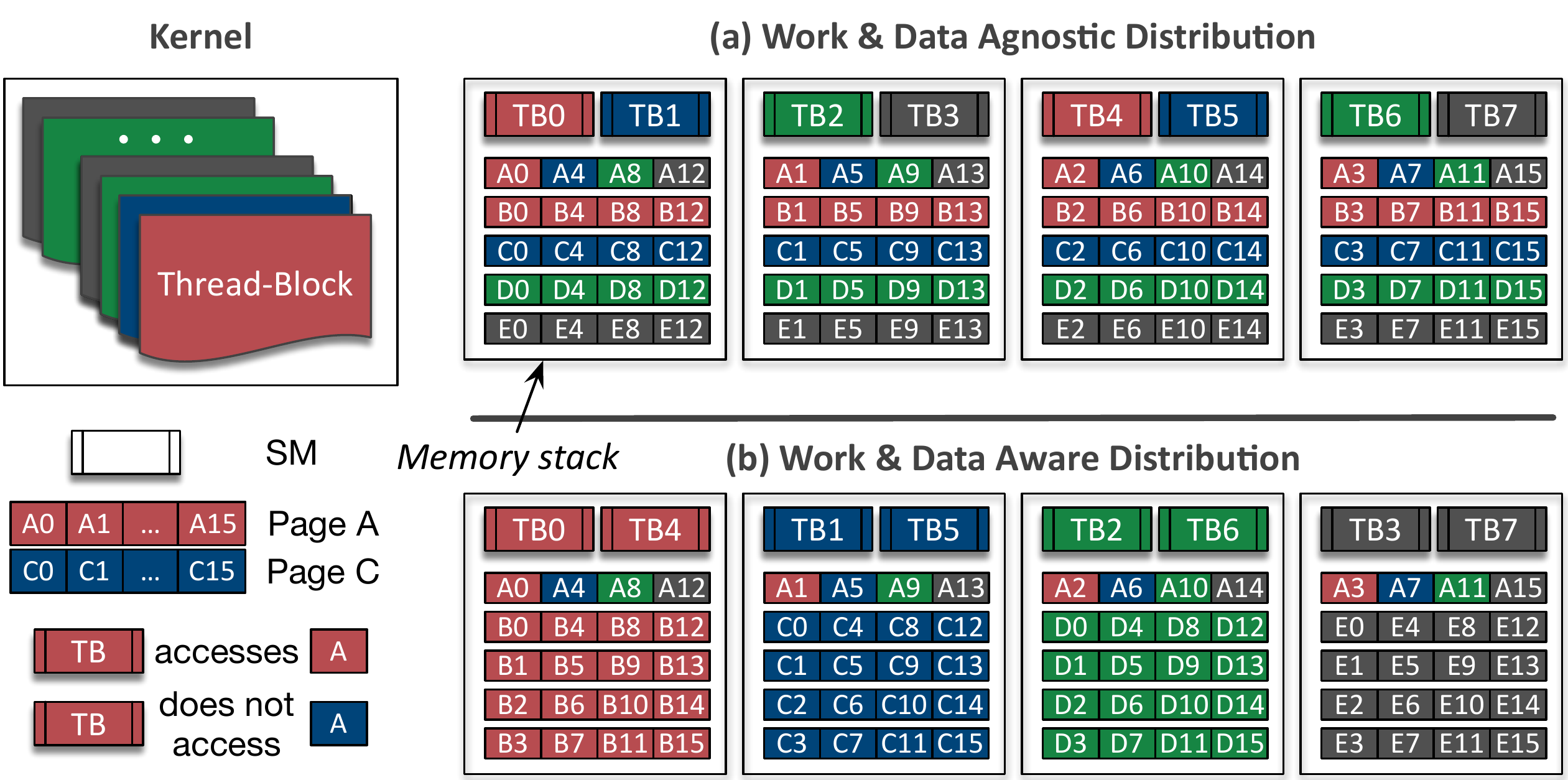}
\caption{(a) shows an example non-NDP-aware distribution of thread-blocks (denoted as \textbf{TB}) and pages (denoted as \textbf{A}, \textbf{B}, \textbf{C}, \textbf{D} and \textbf{E}) across memory stacks. (b) demonstrates how an NDP-aware mechanism can do better.}
\label{fig:mechanism_overview}
\end{figure}

Figure~\ref{fig:mechanism_overview} (a) shows how a non-NDP-aware mechanism places thread-blocks and memory pages across multiple memory stacks.
In this example, there are four memory stacks in the system, each with two SMs in the logic layer.
Five 4KB memory pages (A, B, C, D, E) are allocated and they are distributed across all memory stacks at 256B granularity.
Each 256B chunk is color-coded depending on which thread-block accesses it. 
For instance, A0, A1, A2 and A3 are accessed by TB0 and TB4, and A4, A5, A6 and A7 are accessed by TB1 and TB5.
Note that page B is accessed only by TB0 and TB4, and page C is accessed only by TB1 and TB5.
%Note also that for ease of explanation, we ignore the case where a subpage (or even a cacheline) is accessed by more than one thread-block.
Accesses from TB0 to A0, B0, B4, B8 and B12 are efficient since thread-block and data are located in the same memory stack, while those to the rest (A1, A2, A3, B1, B2, B3, B5, \ldots, B15) are not.

%\vspace{-0.5em}

Figure~\ref{fig:mechanism_overview} (b) demonstrates how an NDP-aware mechanism can place the thread-blocks and memory pages for efficient use of NDP. 
Page B, C, D and E are allocated in different memory stacks and the thread-blocks that access each page exclusively are placed in the corresponding memory stacks.
With this co-location of thread-blocks (computations) and the data they exclusively use, all the accesses to the page B from TB0 and TB4, those to the page C from TB1 and TB5, etc, are efficient, exploiting the large internal memory bandwidth.
Note that page A is still distributed across memory stacks since it is accessed by all thread-blocks (shared).

\subsection{Dual-mode Address Mapping}
\label{subsec:address_interleaving}

\textbf{Hardware Support.}
We propose to use different sets of bits for address mapping for each memory page depending on the anticipated access patterns, allowing the two sets of mappings to co-exist.
The default (fine-grain) address mapping distributes a page across memory stacks (as is done today), and the alternative (coarse-grain) address mapping allocates (or localizes) an entire page in a single memory stack (as is desirable for NDP exclusive data).
We refer to the distributed page as \textbf{FGP} and the localized page as \textbf{CGP}. 
FGP is better suited for the data that is shared among SMs in multiple memory stacks or accessed primarily by the host processor. 
On the other hand, CGP is better suited for the data that is exclusively accessed by the SMs in one memory stack.
Note that once hardware provides the ability to map an entire page to one memory stack (as is enabled by our selective use of coarse-grain address mapping), an NDP-aware operating system (OS) could allocate arbitrarily large objects within one memory stack by mapping all the virtual pages of that object to the physical pages (CGPs) in the memory stack.

\begin{figure}[htb]
\centering
\includegraphics[width=\columnwidth,keepaspectratio]{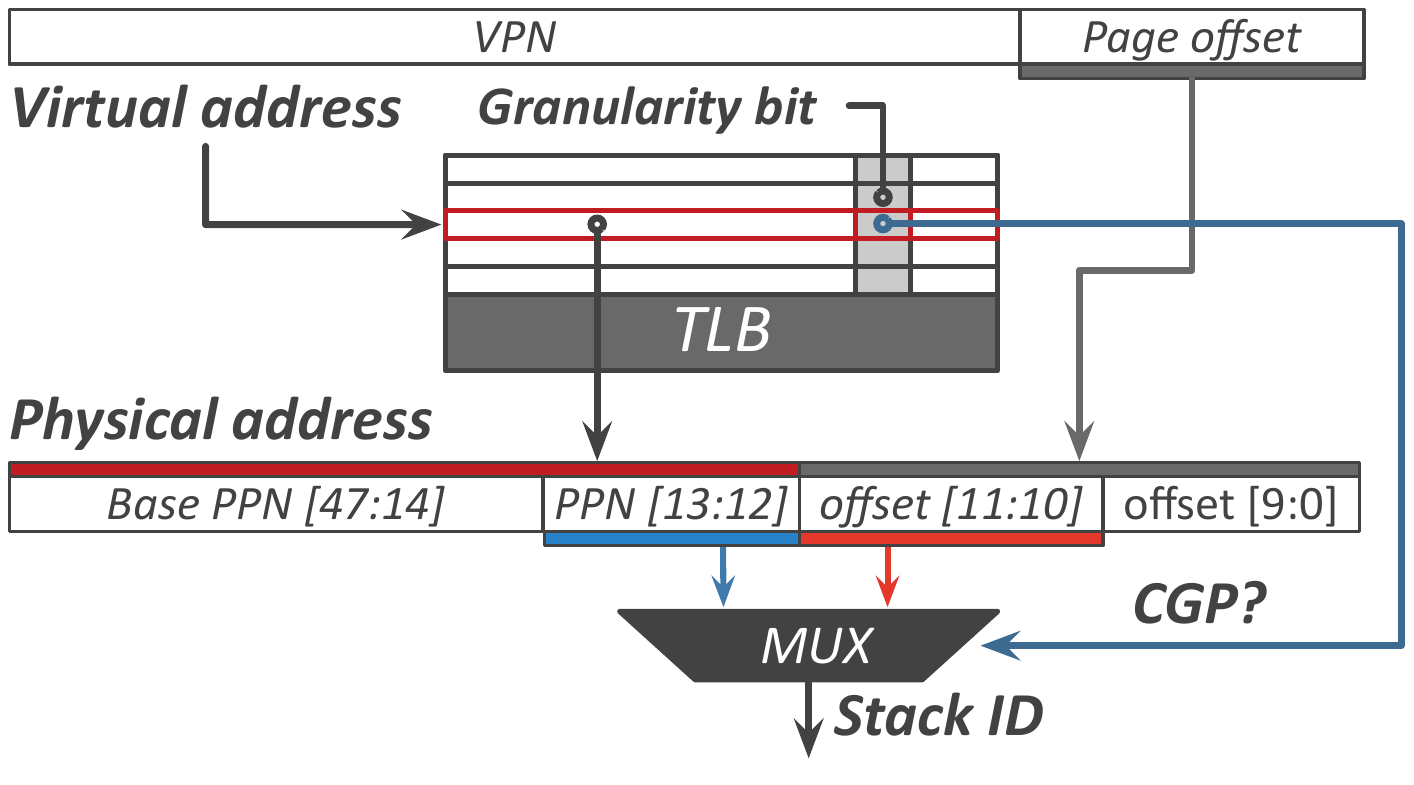}
\caption{Hardware for a dual-mode address mapping}
%\vspace{-1em}
\label{fig:mechanism}
\end{figure}

PTEs, TLB entries and cache lines are extended to indicate the granularity information, fine-grain or coarse-grain, for each page, as shown in Figure~\ref{fig:mechanism}. 
The granularity bit in a PTE is set by the OS when the CGP is allocated, and the granularity bit in a cache line is set when the cache line is allocated.
When the granularity bit is set, indicating CGP, the lowest bits from the PPN (Physical Page Number) are used to index memory stack, whereas the highest bits from the page offset are used for the FGP.
For example, in a system with four memory stacks, when a cache line is evicted from the last level cache, a write-back request is sent to the memory stack indexed by either the bits [13:12] when the granularity bit is set (for the CGP) or the bits [11:10] when the granularity bit is not set (for the FGP).
Be assured that we only change the mapping of the physical address to memory stacks and not the physical address itself.
Thus, cache is accessed with the original physical address, irrespective of the granularity information, and our mechanism does not have any impact on the cache coherence protocol or virtual address translation.

\textbf{System Software Support.}
The OS should be aware of the dual-mode address mapping (1) to indicate the granularity information in the PTEs and TLB entries, and (2) for page management, such as free page management or page replacement. 
It is important to note that it requires a set of adjacent FGPs to allocate a CGP (technically, a set of CGPs are allocated together).
Consider a system where an FGP spans N consecutive memory stacks, occupying a contiguous block of M bytes in each memory stack. 
In that system, a CGP occupies N$\times$M contiguous bytes within a single memory stack. 
Therefore, a single CGP occupies the space that would have been utilized by N different FGPs within one memory stack (but does not utilize any of the space those N FGPs would have occupied in other memory stacks). 
As a result, each block of N contiguous pages must uniformly be configured as FGP or CGP to avoid data layout conflicts. 
However, different blocks of N pages may be independently configured as FGP or CGP based on application or OS requirements.

\begin{figure}[htb]
\includegraphics[width=\columnwidth,keepaspectratio]{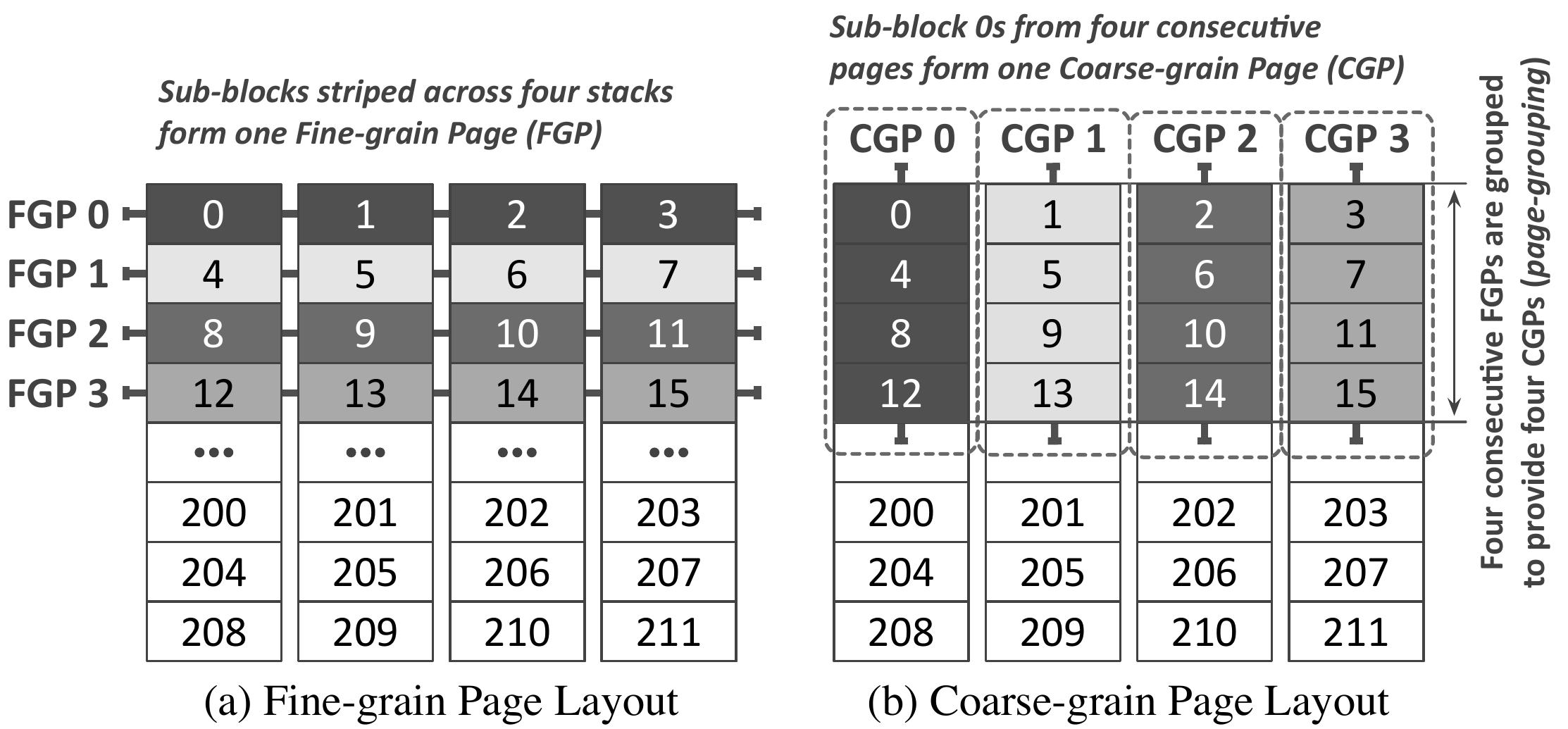}
\caption{Conceptual diagram of page-group. The number indicates a memory block address and the blocks of the same color belong to the same OS page.}
%\vspace{-1em}
\label{fig:fgr_cgr_example}
\end{figure}

For example, when FGP 0 in Figure~\ref{fig:fgr_cgr_example} (a), consisting of block 0, 1, 2 and 3, is converted to a CGP, there are conflicts with block 4, 8 and 12 from the three subsequent FGPs (each from FGP 1, FGP 2 and FGP 3, respectively).
Therefore, those four FGPs must be converted to CGPs together, as shown in Figure~\ref{fig:fgr_cgr_example} (b).
We use the term \textit{page-group} to refer to such a set of pages that must be converted together.
Hence, the OS should decide between FGP and CGP at a page-group granularity and can switch between FGP and CGP only when all the pages in the page-group are free.

\subsection{Compute-Data Co-location Algorithm}
\label{subsec:algorithm}

In traditional GPUs, thread-blocks can be scheduled in any order, as they are supposed to run concurrently. 
The number of thread-blocks that can be run together in one SM is determined by thread-block resource constraints.
Normally, the thread-blocks are scheduled in order and as soon as one thread-block retires, the next thread-block is scheduled to any available SM. 
However, to benefit from careful data placement, as is enabled by our dual-mode address mapping mechanism, thread-blocks and the data they access must be co-located in the same memory stack.
To steer thread-blocks and the data they access to the same memory stack, we set an affinity between thread-blocks and memory stacks.

\subsubsection{Affinity-based Work Scheduling Algorithm}

We compute which memory stack each thread-block has \emph{affinity} to using the following equation.

\begin{dmath}
affinity = \left ( \frac{block\_id}{N_{blocks\_per\_stack}} \right ) \bmod N_{stacks}
\label{eq:bsch}
\end{dmath}

$block\_id$ is flattened for multi-dimensional data based on row-major ordering, i.e., {\tt blockIdx.y} $\times$ {\tt blockDim.x} + {\tt blockIdx.x}.
$N_{blocks\_per\_stack}$ is the number of thread-blocks that can run concurrently in one memory stack. 
For example, if one memory stack has four SMs and each of which can run six thread-blocks, $N_{blocks\_per\_stack}$ is 24. 
When \emph{N} is the number of memory stacks and \emph{T} is the total number of thread-blocks, \nicefrac[]{T}{N} thread-blocks have the same affinity.
With the affinity information, whenever an SM is available, instead of assigning any unscheduled thread-block to it, the scheduler picks one that has affinity to that memory stack.\footnote{This scheduling algorithm is conceptually similar to the guided scheduling policy in OpenMP, where programmer specifies chunk size (the number of loop iterations that one thread executes).}
This may potentially lead to load imbalance compared to the baseline of assigning any available thread-block to any SM in the system.
However, the number of thread-blocks typically being much greater than the number of memory stacks reduces the likelihood of load imbalance. 

The hardware and runtime system must be extended to support this modified scheduling scheme.
The scheduling algorithm could be optimized further to select thread-blocks from other stacks when a memory stack does not have any work left to do, similar to the work-stealing algorithm, for the potential work imbalance that can occur when there are large differences among the amount of work across thread-blocks.
However, in our 20 evaluated benchmarks, only one suffered performance degradation due to the affinity-based scheduling algorithm.
Therefore, we did not implement the work-stealing optimization.

\subsubsection{Data Placement Algorithm}
\label{subsec:data_placement_algorithm}

\begin{figure}[htb]
\centering
\includegraphics[width=\columnwidth,keepaspectratio]{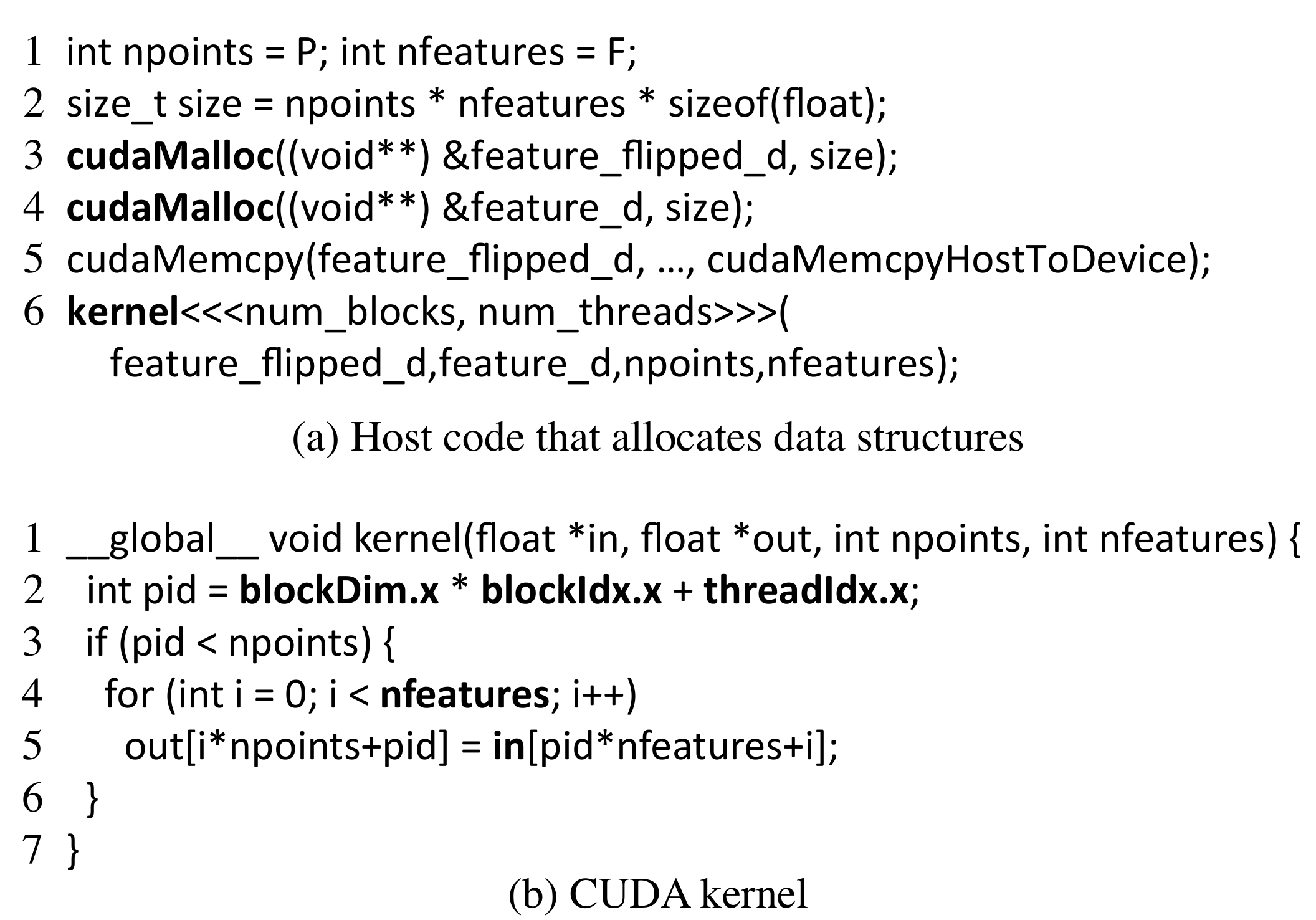}
\caption{Code snippet from K-means Clustering. (a) shows the host code. (b) shows the kernel code.}
\label{fig:kmeans-code}
\end{figure}

While dual-mode address mapping enables the ability to localize an entire page in a single memory stack, the question of how to identify the exclusively accessed or shared pages remains.
This identification is particularly difficult for GPU systems because data structures are allocated by the host processor before the kernel invocation and used by all threads in the kernel later.\footnote{We only target global data structures, which are used by the threads in the system since local data structures are easily identifiable with specific keywords.}
For example, Figure~\ref{fig:kmeans-code} (a) shows the host code that allocates data structures (line 3-4) and initializes them (line 5). 
In the kernel, shown in Figure~\ref{fig:kmeans-code} (b), each thread accesses \texttt{nfeatures} elements (line 4) from \texttt{(pid $\times$ nfeatures)-th} element of \texttt{feature_flipped_d} array (line 5).
Since the amount of data each thread (and thread-block) accesses is \textit{unknown} at the time data structure is allocated, it is not trivial to partition (or place) data appropriately.

\vspace{-0.5em}

To realize an NDP-aware data placement, we propose a compiler- and/or profiler-based technique that identifies the amount of data used by one thread-block for each data structure and decides which address mapping is desirable for the data structure (technically, for the pages in which the data structure is allocated).
It is based on the following four observations.
First, the amount of data used by one thread-block is often determined by the number of threads in a thread-block and the size of data structure that each thread accesses.
Second, compile-time (symbolic) analysis can be used to detect if there exists a regular access pattern for a data structure.
Third, profiler-assisted techniques can be used to estimate input-dependent accesses (more on this is explained later).
Fourth, although the number of threads in a thread-block is often input-dependent, it is determined before a kernel invocation (generally, even before data structures are allocated).

Based on these observations, we implemented the compile-time analysis on LLVM infrastructure.
We extended the FunctionPass, which enables traversing all the kernel functions at compile time, and performed the symbolic analysis.
For all the memory accesses inside the kernel function, we analyzed the ``GetElementPtrInst'' LLVM instruction, which performs the index computation.
Based on the index expression and the types of variables it uses, we examine if there exists a runtime-constant stride between two consecutive thread-blocks.\footnote{In the examination, we check if expression uses only the 1) kernel-invocation-constants, such as parameters, block/grid dimensions or global constants, which are determined before kernel invocation and remain constant throughout the kernel execution, 2) thread index, thread-block index and/or loop index (for local loops in the kernel).}  
If such a stride is found, we insert instructions in the host code to compute the stride distance between two consecutive thread-blocks at runtime.
We use profiler-assisted techniques for the case where the access pattern is input-dependent \emph{and} only when the input is not changed frequently (e.g., graph computing workloads). 
Note that the profiler performs a similar examination as the compile-time analysis.
Also our mechanism use FGP for irregularly accessed data, shared data or parameter objects, as they are accessed by many thread-blocks.

Where the data should be located can also be computed, as the affinity-based work scheduling algorithm already determines where the computation will be performed.
For example, if one thread-block accesses the first $B$ bytes of a memory object and $N$ consecutive thread-blocks will be scheduled to the SMs in a memory stack, the mapping algorithm allocates contiguous chunks of $B \times N$ bytes on each memory stack.
The equations to compute $chunk\_size$ and $stack\_id$ are as follows:

\begin{dmath}
chunk\_size = min(4KB, B \times N_{blocks\_per\_stack})
\label{eq:cs}
\end{dmath}

\begin{dmath}
  stack\_id = \left ( \frac{virtual\_addr -
      obj\_start\_addr}{chunk\_size} \right ) \bmod N_{stacks}
\label{eq:si}
\end{dmath}

Please note that the $chunk\_size$ is upper-bounded by 4KB since an arbitrary number of pages can be allocated in a single memory stack for any large object with hardware support to map an entire page to a single memory stack with CGP.
$obj\_start\_addr$ is the starting virtual address of an object.
When the $chunk\_size$ is not a multiple of physical page size, we round up to the next multiple of pages. 
The resulting misaligned pages will be shared by SMs from two consecutive memory stacks, but this is still better than un-aligned distribution of data across all memory stacks. 
Commonly, $N_{blocks\_per\_stack}$ is moderately big since multiple thread-blocks can run concurrently on an SM, which often results in a big $chunk\_size$ (greater or close to 4KB).
Note that programs often use more than one data structure.
Our proposed mechanism support multiple data structures since we compute the $chunk\_size$ for each data structure using its own $B$ size based on the structure's access pattern.

We demonstrate how our data placement algorithm works with Figure~\ref{fig:kmeans-code}, a code snippet from K-means Clustering.
The size of each data element can be identified (and computed) at compile-time, and the first element and the number of consecutive elements that each thread accesses can also be analyzed with our compile-time analysis routine.
In this example, each thread accesses \texttt{nfeatures} consecutive elements from \texttt{(pid $\times$ nfeatures)-th} element, as shown in lines 4 and 5 of Figure~\ref{fig:kmeans-code} (b). 
Since each thread-block has \texttt{blockDim.x} threads, \texttt{blockDim.x $\times$ nfeatures $\times$ sizeof(float)} is the $B$ value.
This means that the first thread-block accesses $B$ bytes from the starting address of the array (\texttt{in}) and the second thread-block accesses next $B$ bytes.
Note that the number of thread-blocks and threads per thread-block are determined before the kernel invocation.

When a {\tt cudaMalloc} function is called, our extended runtime system uses this information and the $B$ information to compute the $chunk\_size$ using Eq~\eqref{eq:cs} for the corresponding data structure and decides whether it should be allocated with the FGP or CGP. 
If a data structure is accessed by multiple kernels, the information of the first kernel that accesses it is used to compute the number of thread-blocks per memory stack.  
Accesses to 3-D data structures are often more complicated than those to 1-D or 2-D data structures, for which the index is typically computed with both {\tt blockDim.x} and {\tt blockDim.y}.
In this paper, we focus on 2-D data structure and leave the extension to support the 3-D data structures and more complex data structures for the future work.  

\section{Evaluation Methodology}
\label{sec:method}
\subsection{Hardware Configurations}
\label{subsec:hardware} 

%\vspace{-1.75em}
\begin{table}[htb]
\caption{Evaluated system}
\label{table:system_configuration}
\includegraphics[width=\figwidth,keepaspectratio]{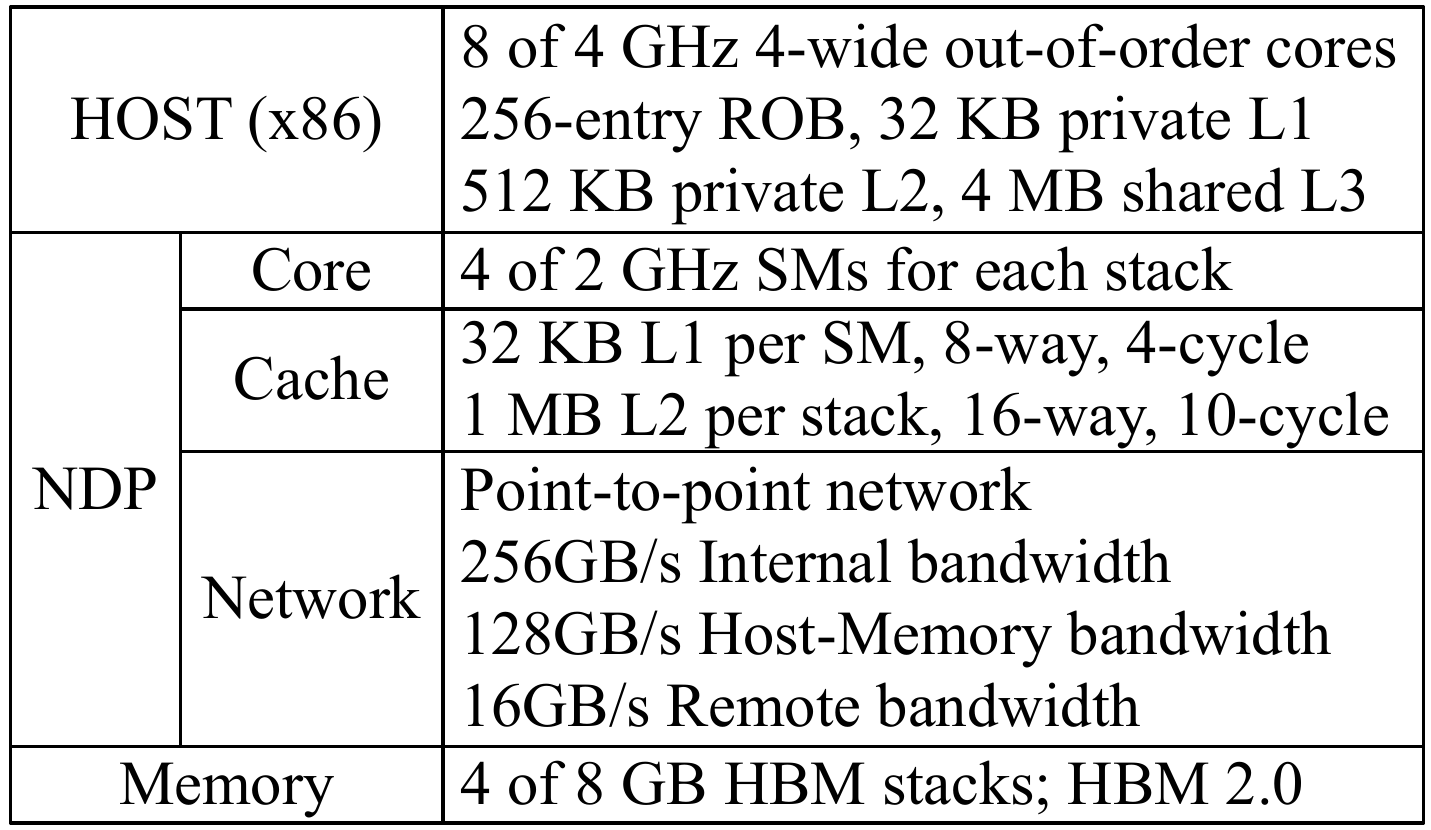}
\centering
\end{table}

\begin{table*}[htb]
\caption{Benchmark categories}
\label{table:benchmark}
\includegraphics[width=\textwidth,keepaspectratio]{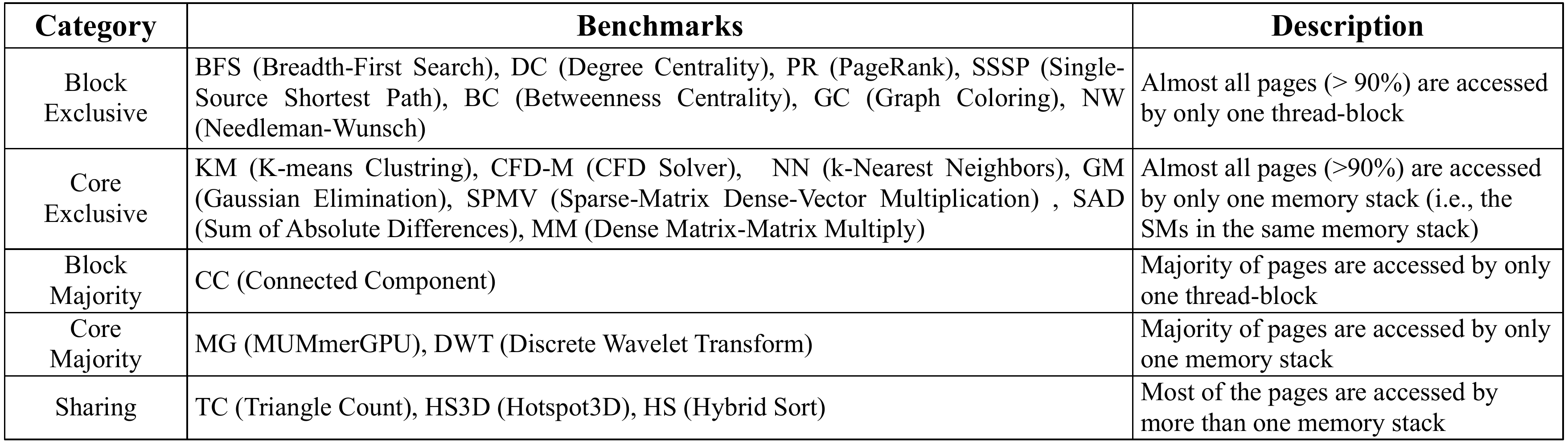}
\end{table*}

We evaluate our mechanism using SST~\cite{rod:hem11} with MacSim~\cite{ros:cop11}, a cycle-level microarchitecture simulator. 
Low-level DRAM timing constraints are faithfully simulated using DRAMSim2~\cite{ros:cop11}, which was modified to model the HBM 2.0 specification~\cite{hbm2}. 
Our default system configuration comprises the host processor and four HBM-based memory stacks, where each memory stack consists of four SMs and 8GB HBM memory. 
More details on the simulated system configuration are provided in Table~\ref{table:system_configuration}.
We use 128-byte interleaving and 4KB interleaving to form the FGR and CGR, respectively.
Each channel is modeled to provide 32 GB/s of peak memory bandwidth; therefore 256 GB/s of total internal memory bandwidth is exploitable by the SMs in the logic layer.
We assume 128 GB/s of aggregate memory bandwidth is available for the Host network.
We model a Remote network to provide 16 GB/s of memory bandwidth.
We also perform detailed sensitivity studies, where we vary the bandwidth of Local, Host and Remote network.

\subsection{Benchmarks}
\label{subsec:bench}

We use 20 benchmarks from GraphBIG~\cite{lif:yin15}, Rodinia~\cite{che:boy09}, and Parboil~\cite{parboil}.
We classify a benchmark as being \textbf{block-exclusive} if almost all pages (> 90\%) are accessed by only one thread-block, \textbf{core-exclusive} if almost all pages (> 90\%) are accessed by one memory stack (i.e., multiple SMs in the same memory stack), \textbf{block-majority} if the majority of pages (> 60\%) are accessed by only one thread-block, \textbf{core-majority} if the majority of pages (> 60\%) are accessed by one memory stack, and \textbf{sharing} if most of the pages are accessed by more than one memory stack.
Table~\ref{table:benchmark} summarizes the benchmarks and the category they belong to.

\section{Evaluation Results}
\label{sec:eval}

\subsection{Performance}
\label{subsec:eval:perf}

%\vspace{-0.75em}
\begin{figure}[!htb]
\includegraphics[width=\figwidth,keepaspectratio,angle=0]{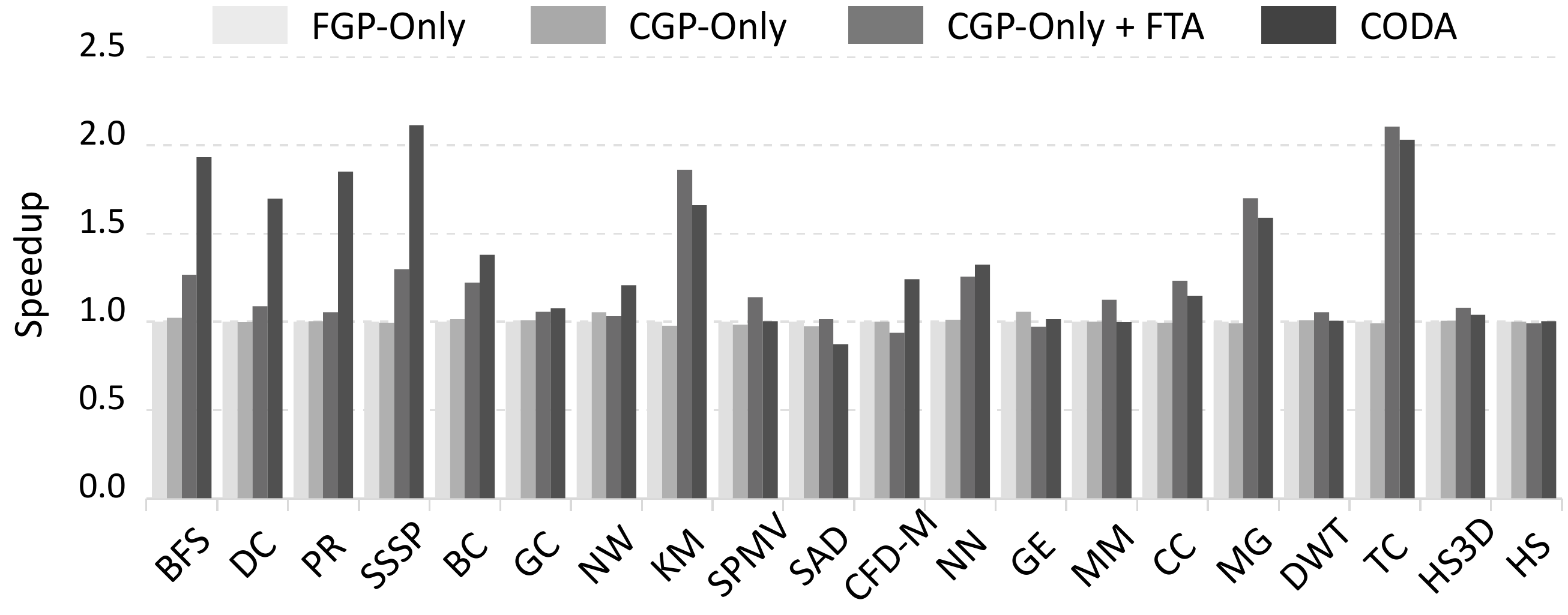}
\caption{Speedup of {\mech} over FGP-Only, CGP-Only, and an idealized first-touch-based allocation scheme (CGP-Only + FTA)}
\label{fig:eval_perf}
\end{figure}
%\vspace{-0.5em}

Figure~\ref{fig:eval_perf} shows the performance improvement of {\mech} for the benchmarks described in Table~\ref{table:benchmark}.
FGP-Only represents the baseline where every page is distributed across memory stacks at 128-byte fine-grain interleaving granularity, and CGP-Only represents the case where consecutive 4KB pages are allocated in consecutive memory stacks in a circular order; this represents affinity-unaware data placement even when coarse-grain data allocation is available. 
CGP-Only+FTA (First-Touch-based Allocation)\footnote{Here, first-touch-based allocation  scheme places each physical page on the memory stack that first accessed it. We ignore the accesses by the host processor for the purpose for determining the first access since all the pages are initially allocated by the host processor before the kernel invocation.} represents the case where each page is allocated to the memory stack that first touches the page. 
Even though this is not a practical implementation due to the lack of first-touch information at the time data is allocated (and often initialized) by the host processor, this can be a good indicator of the potential effectiveness of coarse-grain allocation for each benchmark.\footnote{One simple way to implement first-touch-based allocation is to migrate pages upon first access. We observed that this migration-based first-touch allocation is not very effective (not shown, 7\% speedup, as opposed to 19\% speedup of CGP-Only+FTA) mainly due to small number of reuses of memory pages after migrations (due to burst and clustered access patterns); that is, the migration overhead is not mitigated. This makes a case for better data allocation rather than reactive data movement.}

Our evaluation results show that {\mech}~outperforms both FGP-Only and CGP-Only by 31\%.
{\mech}~even outperforms CGP-Only+FTA for most benchmarks.
Allocating an entire page in the memory stack that exclusively accesses it if the page is exclusively accessed by one memory stack brings a substantial reduction in remote data accesses and increase in local data accesses.
%In addition, distributing a page across all memory stacks if the page is accessed by SMs in more than one memory stack reduces the memory bandwidth contention and its negative consequences.
This variation in remote and local data accesses directly leads to the performance improvement, as remote data accesses are limited by the low bandwidth of the off-chip links, whereas local data accesses exploit the large internal memory bandwidth.
Perhaps more importantly, such bandwidth discrepancy becomes even more pronounced as the interconnection network becomes overwhelmed with more remote data accesses.
Though lower bandwidth of the off-chip links does not necessarily mean longer memory access latency, when coupled with the off-chip communication overheads such as queuing delays and/or external transfer time, average memory access latency can be significantly affected by the number of remote data accesses as well. 

Our mechanism achieves 1.56x and 1.13x average performance improvements over the baseline for \textbf{block-exclusive} and \textbf{core-exclusive} benchmarks, respectively.
This is particularly effective in graph algorithms with large numbers of neighbor accesses (e.g., \texttt{BFS}, \texttt{DC}, \texttt{PR}, and \texttt{SSSP}), which are difficult to handle efficiently without an NDP-aware data allocation.
Even for the \textbf{sharing} benchmarks in which most pages are accessed by many SMs, our mechanism can localize accesses whenever possible and achieves 1.29x average performance improvements over the baseline.

\subsection{Local vs Remote Access}
\label{subsec:eval:remote}

%\vspace{-0.75em}
\begin{figure}[!htb]
\includegraphics[width=\figwidth,keepaspectratio,angle=0]{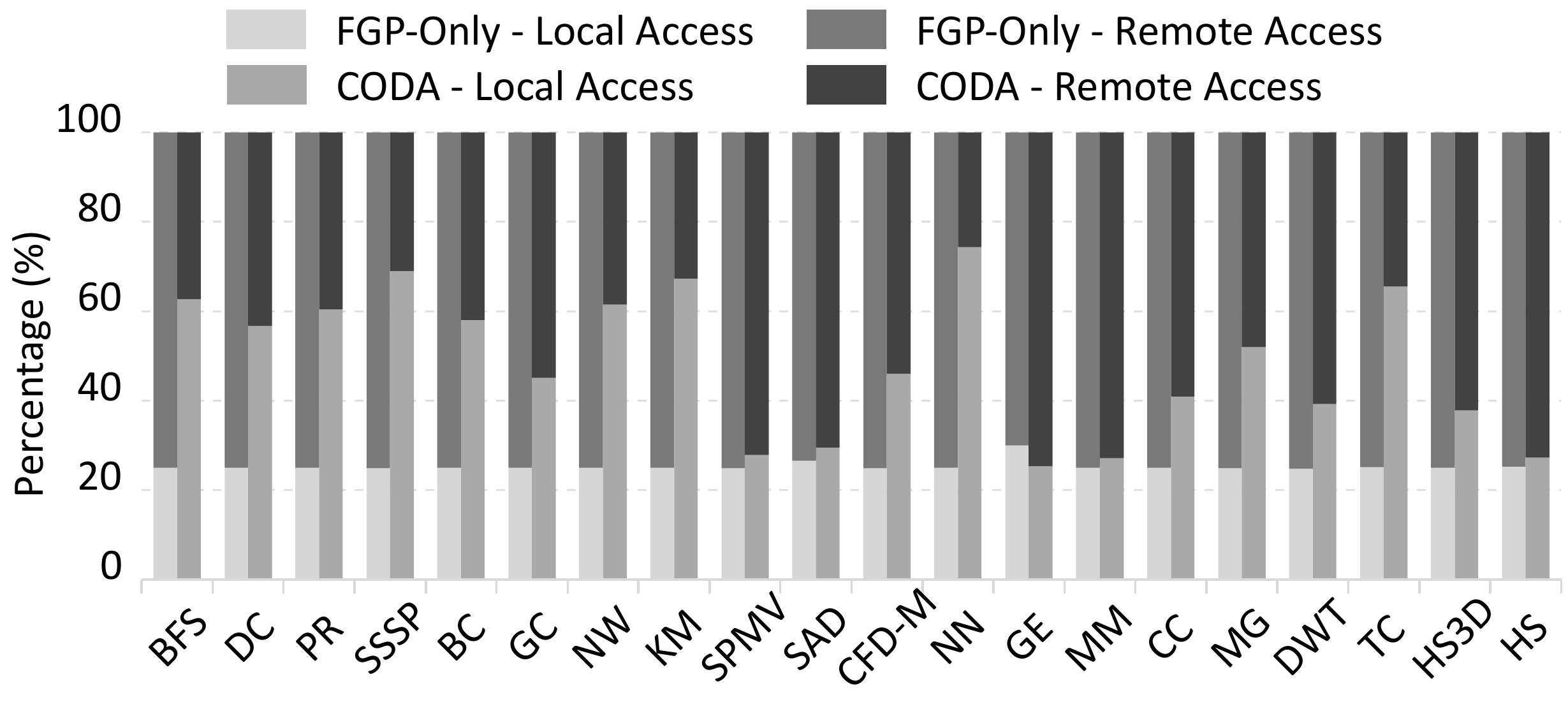}
\caption{Comparison of local and remote data accesses between FGP-Only and our mechanism (CODA)}
\label{fig:eval_acc_var}
\end{figure}

Figure~\ref{fig:eval_acc_var} shows distribution of memory accesses, local versus remote, for the baseline and how it varies with our mechanism.
Our mechanism significantly reduces remote data accesses for all the evaluated benchmarks but one, \texttt{GE}.
A substantial reduction in remote data accesses and an increase in local data accesses contribute to the performance improvement for the following reasons.
First, local data accesses can utilize the large internal memory bandwidth, while remote data accesses are limited by the lower memory bandwidth of the off-chip links.
Second, for the remote data accesses, a great amount of time could be spent on waiting for network due to the off-chip communication.
This can be incurred as a result of limited network bandwidth, but can be exacerbated further due to the artifacts of the off-chip communication, such as queuing delays, routing delays, etc.
Our mechanism significantly reduces remote data accesses, enabling the utilization of large internal memory bandwidth and also mitigating the effect of interconnection network congestion by placing data objects in the same memory stack in which the computation is to be performed.

Our mechanism is especially effective for the block-exclu\-sive and core-exclusive benchmarks.
On average, 47\% and 34\% remote data accesses are reduced, respectively.
Even for the sharing benchmarks, by identifying the pages that are accessed by a few thread-blocks or SMs and allocating them where the computation is to be performed, our mechanism reduces 32\% remote data accesses.

%\subsection{Transfer Energy}
%\label{subsec:eval:energy}
%
%\begin{figure}[!ht]
%\missingfigure[figwidth=\figwidth]{transfer energy reduction}
%%\includegraphics[width=\figwidth,keepaspectratio,angle =0]{figs/eval_transfer_energy}
%\caption{Transfer energy reduction}
%\label{fig:eval_transfer_energy}
%\end{figure}

%Figure~\ref{fig:eval_transfer_energy} shows the normalized data transfer energy consumption. 
%Our mechanism consumes xx\% less average energy compared to the baseline, mainly due to a significant reduction in remote accesses. 
%
%When there are less performance differences between CGR and FGR, CGR can still reduce the remote accesses, which can reduce the energy consumption.  
%Figure~\ref{fig:eval_transfer_energy} shows the energy reduction in data movement cost. 
%FGR, CGR-RR, \mech are shown in this figure.  
%Each bar in this figure is energy savings for the same bar in Figure~\ref{fig:eval_bw_lat_sensitivity}.
%Hence, even when CGR does not provide any additional performance benefit, for the energy benefits, CGR is a good choice for many benchmarks.  

\subsection{Sensitivity to Bandwidth}
\label{subsec:eval:bw}

%\vspace{-0.75em}
\begin{figure}[!ht]
\includegraphics[width=\figwidth,keepaspectratio,angle=0]{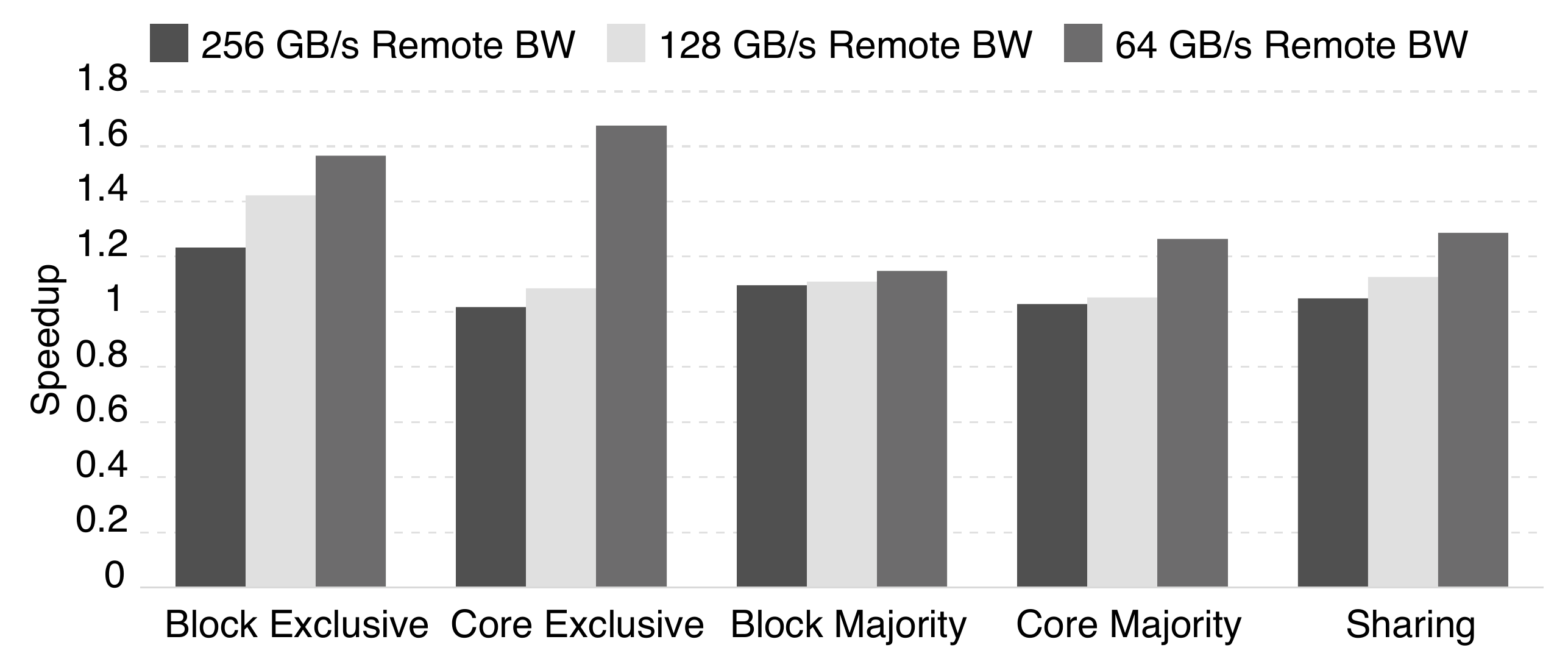}
\caption{Speedup with different remote bandwidth among memory stacks.}
\label{fig:eval_sensitivity_bw}
\end{figure}

Even for highly provisioned systems with unrealistically large Remote bandwidth and low remote memory access latency, co-location of thread-blocks and the data they access improves performance, as shown in Figure~\ref{fig:eval_sensitivity_bw}. 
This is because even in such systems, remote memory accesses cannot be completely free from all resource conflicts.
Careful data placement, as is enabled by our mechanism, can significantly reduce the possibility of such conflicts and therefore can contribute to the performance improvement.

Even when a system has 256 GB/s of aggregated Remote bandwidth, our mechanism improves performance by 8\% (up to 23\%).
It should be noticed that as the gap between Local bandwidth and Remote bandwidth increases (Remote bandwidth is decreased while Local bandwidth remains the same), our mechanism provides more benefit by reducing remote data accesses and opening up more opportunity to exploit large internal memory bandwidth, thereby mitigating the performance penalty of the off-chip communication (performance improvement goes up to 15.2\% and 37.4\%, respectively).

\subsection{Sensitivity to Graph Properties}

In graph computing, the number of vertices and their neighbors that each thread-block accesses depends highly on graph properties. 
To examine the impact of the graph properties on our proposed mechanism, we differentiate the properties that can be estimated at the time the graph is preprocessed\footnote{The term preprocessing generally implies a heavy-weight operation such as a clever partitioning to reduce communication. In this study, however, we only extract basic properties of a graph without scanning through the entire graph.} from those that cannot be estimated.
Basic graph properties such as the number of vertices and edges can be obtained at the time the graph is preprocessed. 
These, combined with the number of threads per thread-block, which is determined based on the resource constraints of the underlying hardware, can be used to estimate the average number of edges that each thread-block accesses ($\mu$) before a kernel invocation and the standard deviation ($\sigma$) of it.
The coefficient of variation of a graph, which can be estimated as \nicefrac[]{$\sigma$}{$\mu$}, is a good indicator of how regular a graph is: graphs with a smaller coefficient of variation is regular.
Therefore, the granularity at which the graph should be distributed, or the block stride distance, can be determined.

\begin{figure}[!ht]
\includegraphics[width=\figwidth,keepaspectratio]{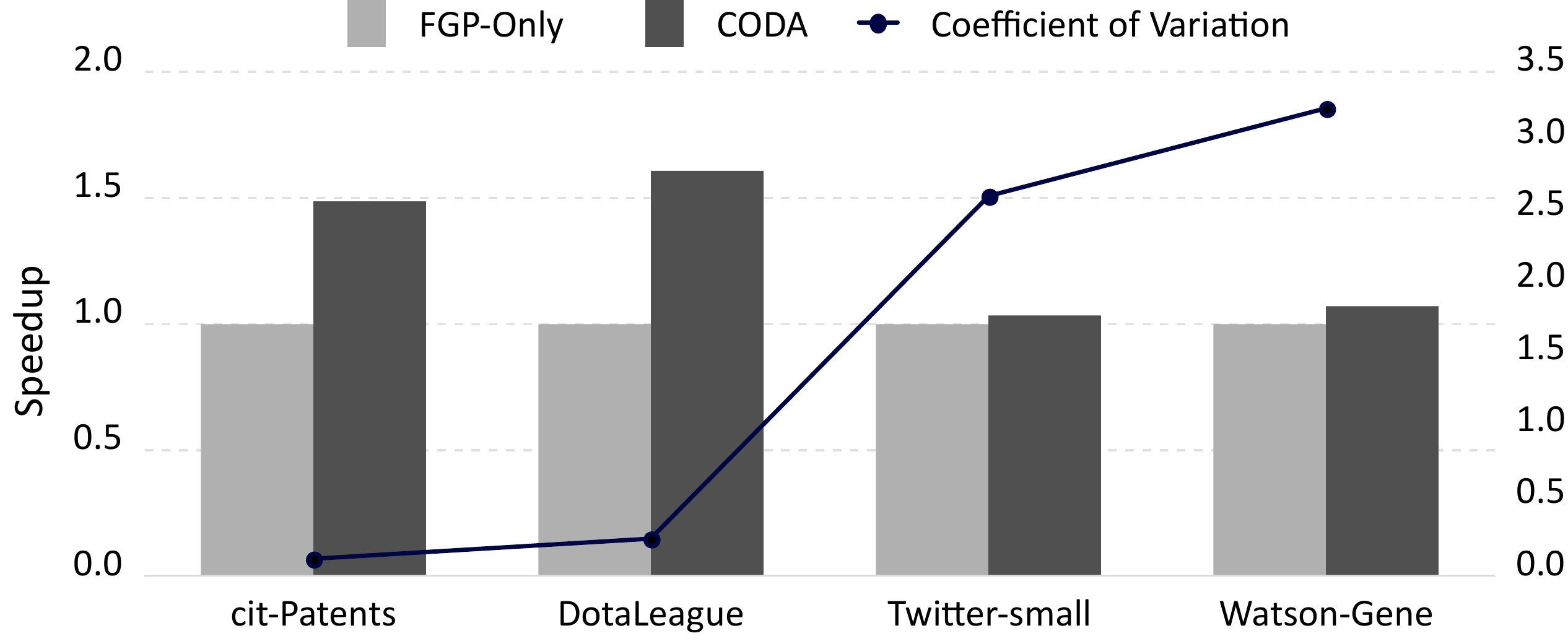}
\caption{PageRank performance with different graphs}
\label{fig:eval_input_sensitivity}
\end{figure}
%\vspace{-0.5em}

Figure~\ref{fig:eval_input_sensitivity} compares the performance of FGP-Only and {\mech}, using the PageRank workload.
The evaluation is based on four real-world graphs, which have 59K to 9M vertices.
Graphs are sorted based on their regularity: graphs with a smaller coefficient of variation appear toward the left side of the figure. 
The coefficient of variation of each graph are also depicted.
We confirm that the effectiveness of our mechanism depends highly on graph properties.
Regular graphs benefit more from our mechanism (55\%) than irregular graphs (5\%) since the estimation accuracy depends only on the properties that can be estimated at the time graph is preprocessed. 
Notably, {\mech} does \emph{not} degrade performance in any case since it detects the data objects that are exclusively accessed by one memory stack and localizes them with CGP, while distributing other data objects with FGP, as in the case of FGP-Only.

\subsection{Multiprogrammed Workloads}
\label{subsec:multi}

%\vspace{-0.75em}
\begin{figure}[!ht]
\includegraphics[width=\figwidth,keepaspectratio,angle =0]{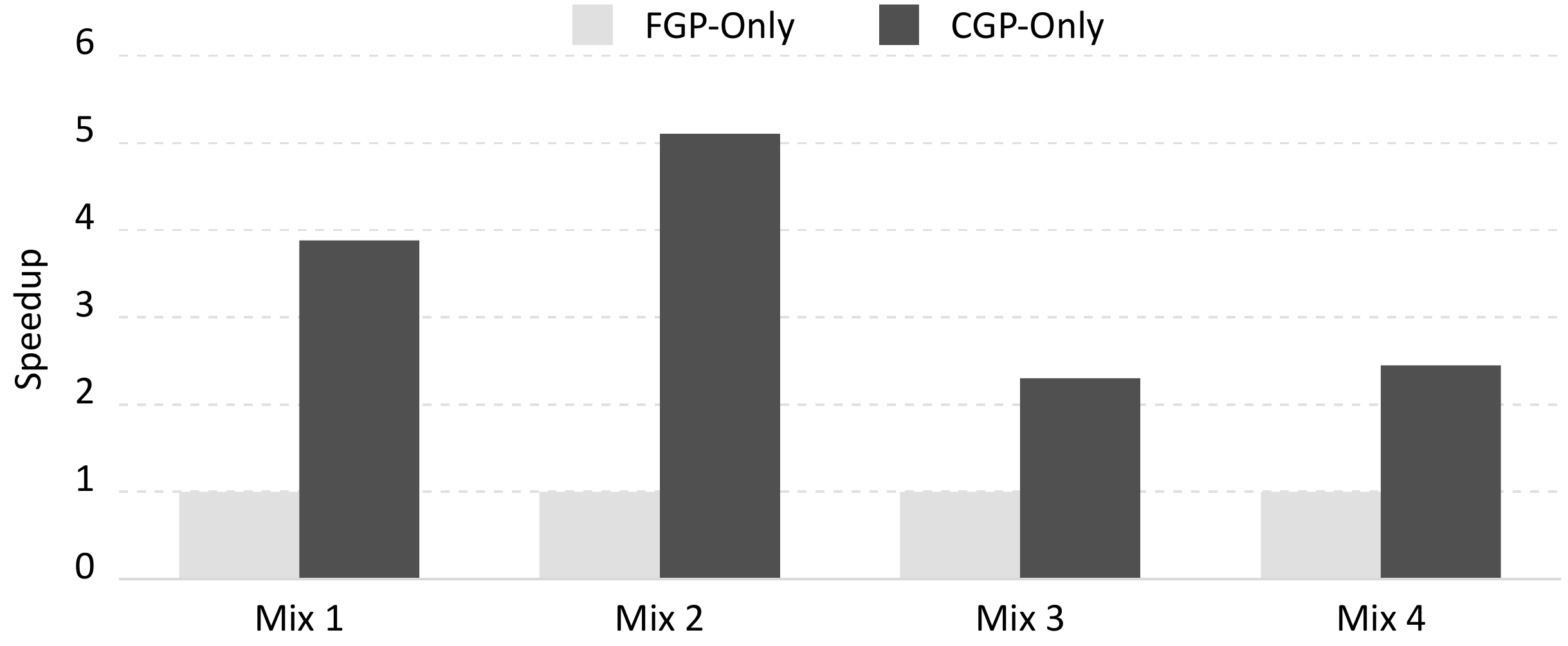}
\caption{Performance of multiple applications}
\label{fig:eval_multi}
\end{figure}
%\vspace{-0.5em}

To further analyze the impact of having hardware that provides the ability to map an entire page to a single memory stack using CGP, we evaluate our CGP-Only configuration with four mixes of multiprogrammed workloads.
Each benchmark is chosen randomly from each category to construct a multiprogrammed workload.
Figure~\ref{fig:eval_multi} compares the performance of CGP-Only with that of FGP-Only, showing that the CGP-Only outperforms the FGP-Only for all the workloads.
With FGP-Only hardware, every memory page is distributed across all memory stacks, which results in a significant number of remote data accesses from all applications.
With hardware that can map an entire page to a single memory stack, as enabled by our mechanism, however, memory pages that an application accesses can be allocated to the memory stack where the application is executed, and hence, all the accesses can exploit the large internal memory bandwidth within the memory stack.
This is an important contribution since it is infeasible or difficult to reduce remote data accesses in the presence of multiple workloads running in a system.

\subsection{Impact of Interleaving Granularity}
\label{subsec:eval:host}

%\vspace{-0.75em}
\begin{figure}[!ht]
\includegraphics[width=\figwidth,keepaspectratio,angle =0]{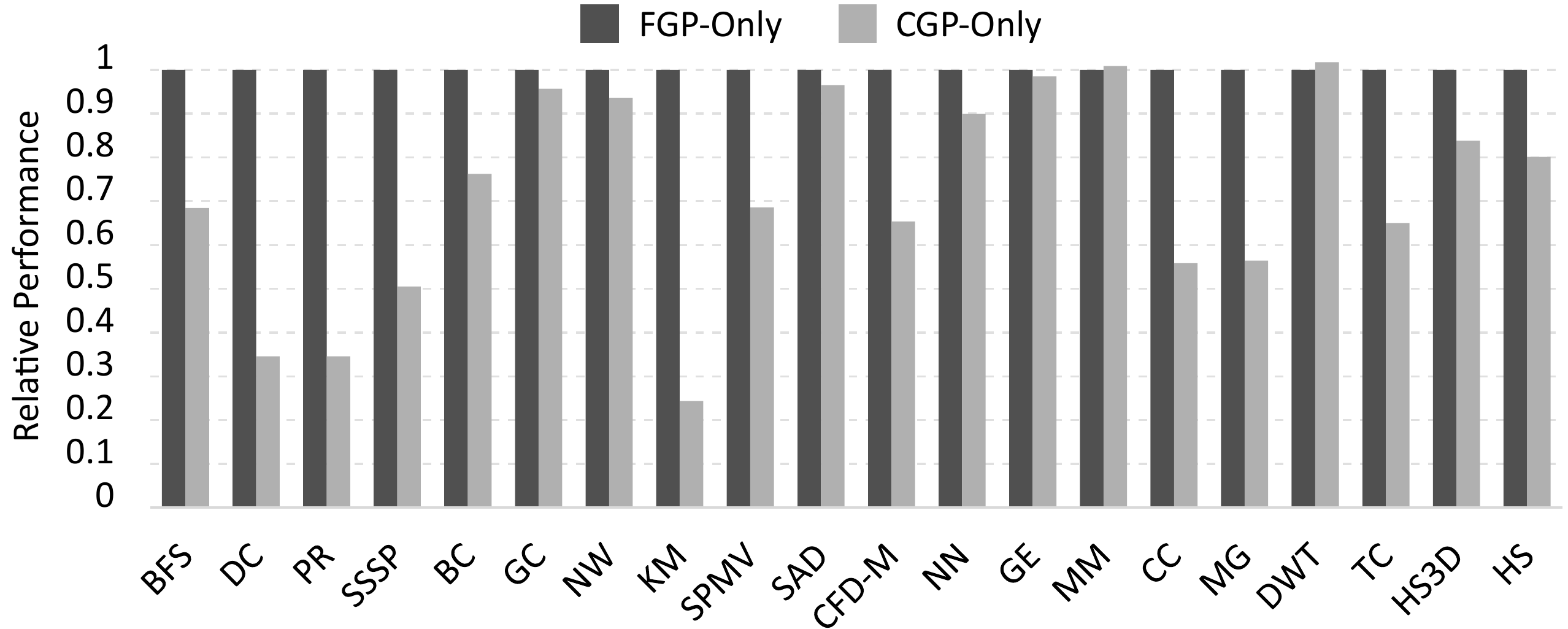}
\caption{Performance impact of interleaving granularity on the host processor}
\label{fig:eval_host}
\end{figure}
%\vspace{-0.5em}

So far we have demonstrated the necessity of the coarse-grain interleaving (technically, selective use of CGP and FGP) for the efficient use of NDP. 
One might consider using \textit{just} coarse-grain interleaving in a system with multiple NDP memory stacks.
However, in this section we present the performance of FGP-Only and CGP-Only for the host side execution (assuming the same overall computational capability as the all NDP stacks) to demonstrate the necessity of the fine-grain interleaving as well.
When an application is run on the host processor, it is desirable that the memory objects it accesses are distributed across multiple memory stacks to achieve maximum memory bandwidth utilization by distributing concurrent accesses across all available memory interfaces.
Figure~\ref{fig:eval_host} shows the performance of the host processor with memories interleaved at different granularities.
FGP-Only and CGP-Only indicate the use of fine-grained interleaved memory and coarse-grained interleaved memory, respectively.
Our evaluation results show that FGP-Only outperforms the CGP-Only by 1.48x due to better memory bandwidth utilization.

\subsection{Impact of Affinity-based Scheduling}
\label{subsec:eval:b2cfix}

%\vspace{-0.75em}
\begin{figure}[!htb]
\includegraphics[width=\figwidth,keepaspectratio]{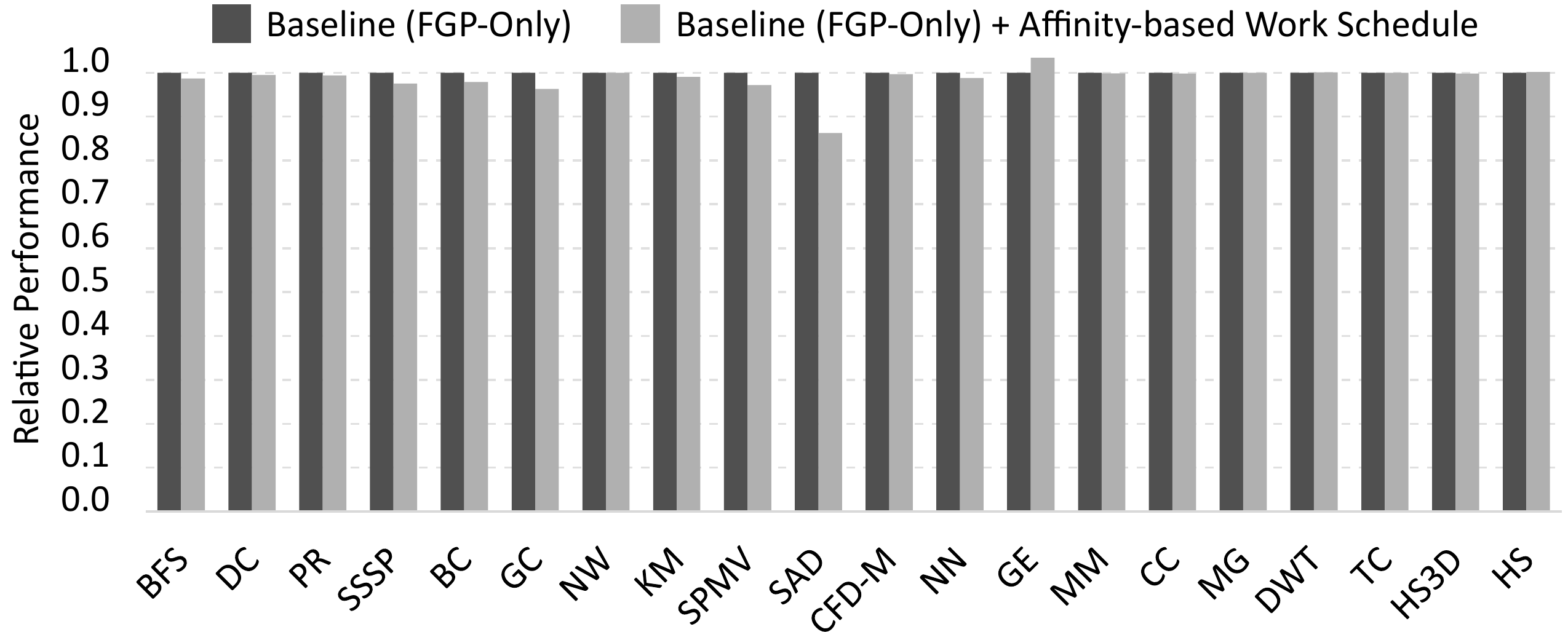}
\caption{Performance impact of an affinity-based work scheduling mechanism}
\label{fig:eval_b2cfix}
\end{figure}
%\vspace{-0.5em}

Thread-blocks cannot be scheduled to any SM with our affinity-based work scheduling mechanism.
In this section, we evaluate the performance impact of the affinity-based work scheduling mechanism. 
Figure~\ref{fig:eval_b2cfix} compares the performance of the affinity-based work scheduling mechanism (FGP-Only + Affinity-based Work Schedule) and that of the baseline (FGP-Only). 
All our evaluated benchmarks are virtually unaffected by the restricted scheduling mechanism, as expected, except for one benchmark, \texttt{SAD}.
The reason why the performance of \texttt{SAD} is degraded by the affinity-based work scheduling is that the number of thread-blocks is small (61) compared to the number of memory stacks and available SMs (16). 
Maintaining load balancing across all available compute resources might be more crucial than carefully co-locating thread-blocks and the data they access, when compute resource bounds the overall performance.
This problem can be alleviated with resource-monitoring-based schemes.
%\vspace{-0.5em}
\section{Discussions}
\label{sec:dis}

\subsection{Complex Address Mapping}
\label{subsec:complex_address_mapping}
 
So far we have assumed a simple address mapping scheme for ease of explanation.
Modern processors, however, use more complex address mapping schemes such as XORing multiple bits (not necessarily consecutive) for channel selection~\cite{pet:dan16}. 
In this section, we discuss the applicability of our dual-mode address mapping mechanism in such systems.
Note that computation and data co-location algorithm presented in Section~\ref{subsec:algorithm} is orthogonal to the address mapping scheme used in the underlying system.
Although the detailed address mapping scheme differs for different architectures, the mappings can be classified into those that use the channel-selection bits exclusively (i.e., they are not used as part of the row- or column-selection), and those that do not (i.e., at least one bit from the channel-selection bits is used as part of the row- or column-selection). 
Our dual-mode address mapping mechanism can be easily extended to support a system with the former class of mappings, where channel-selection bits are used exclusively, by swapping the channel-selection bits with other higher order bits after XOR operation.
However, it is not trivial to support a system with the latter class of mappings, where channel-selection bits are not exclusively used.
One way might be to identify which bits are used exclusively for the channel-selection and which bits are not, and then carefully swapping the channel-selection bits with those that are not used for channel-selection.
This requires further investigation and is a part of our future work.

%\vspace{-0.25em}
\subsection{Large Page and Memory Management}
\label{subsec:large_page}

Large pages have been used to mitigate address translation overheads by reducing the number of PTEs to maintain and increasing TLB hit rates.
However, it comes at a cost, such as an increase in fragmentation and memory footprint.
In this section, we discuss the applicability of our dual-mode address mapping mechanism for the large pages.
Again, computation and data co-location algorithm presented in Section~\ref{subsec:algorithm} is orthogonal to the page size.
First, our dual-mode address mapping can be easily extended for the large pages.
For 2MB pages, for example, address bits [22:21] can be used (instead of address bits [13:12] in the case of 4KB page) to index memory stacks to allocate the entire page in a single memory stack.
However, the key challenge in supporting large page is not about choosing which bits to use for stack selection but about dealing with fragmentation issues.
%To the best of our knowledge, modern GPU systems do not support large pages.
Although our mechanism may complicate page management and potentially increase fragmentation issues, we believe that if page-groups are small (e.g., 4 or 8 pages), this is likely to not be significantly more complicated than normal page management. 
Also, the memory manager can be modified to deal with page-groups for most operations (e.g., flushing out to disk) for better memory management. 
This requires further exploration and is a part of our future work.

\subsection{PTE Extension} 
Our proposed mechanism requires a modification to the PTE format. 
X86 ISA reserves 3 bits [11:9] for future usage~\cite{intel_sw}, so we can use one of the bits to indicate the granularity information. 
When a system employs large pages, extra bits are available in the PTE, which gives more freedom to modify PTE contents. 

%\vspace{-0.25em}
\subsection{NUMA or NUCA Systems}
In this section, we discuss the difference and uniqueness of our system from the conventional NUMA (Non-Uniform Memory Architectures)~\cite{cha:dev94} or NUCA (Non-Uniform Cache Access)~\cite{har:fer09} systems.
First, in NUMA systems, memory policies such as node-local or interleave can be specified and (relatively) easily controlled.
For example, the first-touch based page allocation has already been used in NUMA systems. 
On the contrary, the first-touch based page allocation \textit{cannot} be used in our system due to the lack of first-touch information (recall that data structures are generally allocated and initialized by the host processor before the kernel invocation). 
Even if the first-touch information \textit{was} available, a memory page could not be allocated in a single memory stack without hardware support for the localization.
Moreover, since multiple memory stacks behave like NUMA for GPUs in memory stacks but behave like UMA (Uniform Memory Architecture) for the host processor, we need a mechanism to accommodate the needs of two different computing units.
Second, NUCA systems (e.g., R-NUCA~\cite{har:fer09}) rely on data migration after an access pattern is identified.
The migration overhead is much smaller in NUCA systems than in our system because the former migrates data within a single device (i.e., a tiled L2 cache architecture), whereas the latter migrates data across multiple devices connected via low-bandwidth, high-latency interconnect links.

%\vspace{-0.25em}
\subsection{Other NDP Systems}
The dual-mode address mapping mechanism works irrespective of the type of processing units: CPU, GPU, etc.  
Only the chunk size detection algorithm (in Section~\ref{subsec:algorithm}) needs to be adjusted depending on the programming model.  
Our proposed mechanism will also work well on NDP with conventional DRAM devices~\cite{far:ahn15, bat:sum16}. 
Processing cores in DDRx DIMMs benefit from accessing data in the same DIMM, but the host processor should utilize all DIMMs concurrently. 
Therefore, our proposed mechanism can be very effective to provide such an access localization solution without jeopardizing the host processor's performance.  

%\vspace{-0.75em}
\section{Related Work}
\label{sec:related}

\noindent\textbf{Processing in memory.} 
Processing in memory was proposed decades ago~\cite{gok:hol95, hal:kog99, osk:cho98, pat:and97, mur:kog00, pat:and97-2, kan:hua99}.
Recent advances in 3D stacking technology have given a boost to NDP research~\cite{ahn:yoo15, ahn:hon15, aki:fra15, zha:jay13, zha:jay14, chu:jay13, hsi:ebr16, nai:had17, nai:ant15} to accelerate workloads in various domains (e.g., large-scale graph processing workloads~\cite{ahn:hon15, nai:had17}, Map-Reduce workloads~\cite{pug:jes14}, and HPC applications~\cite{zha:jay14}).
Among these works, Hsieh et al.~\cite{hsi:ebr16} (TOM) addressed the issue of local and remote memory accesses in a system with multiple NDP memory stacks. 
It performs runtime profiling to learn best address mapping for data accessed by offloading candidates, and distributes that data with the discovered mapping. 
In contrast to our proposal, this work 1) essentially delays and decelerates the regular kernel execution because it tests all different address mappings (10 mappings, sweeping from bit position 7 to bit position 16) for all the data accessed by offloading candidates during the runtime learning phase, 2) implicitly assumes a hardware mechanism to distribute data with different mappings. 

%\vspace{0.40em}
\noindent\textbf{Increasing Memory-Level Parallelism.} 
Zurawski et al.~\cite{jur:mur95} presented an address bit swapping scheme to increase memory-level parallelism by reducing the row buffer conflicts in traditional DRAM systems, which is used in AlphaStation 600 5-series workstations.
Zhang et al.~\cite{zha:zhu00} proposed a permutation-based page interleaving scheme in order to reduce row-buffer conflicts and to exploit data access locality in the row-buffer. 
Ghasempour et al.~\cite{gha:jal16} proposed a hardware mechanism to dynamically change the address mapping to increase bank-level parallelism at the cost of a significant amount of page migration overhead.
While our proposed mechanism also uses address bit swapping scheme, it is different from these works in two ways.
First, our mechanism applies address mapping scheme at a page granularity such that pages with different address mappings co-exist in the same memory space. 
Our mechanism is lightweight in a sense that it incurs negligible performance overhead and does not have any impact on the cache coherence protocol or virtual address translation.
Second, our mechanism does not require large-scale page migrations; only a few (e.g., four or eight, depending on the number of memory stacks) pages are affected, since we selectively use CGP at the page-group granularity. 

%\vspace{0.40em}
\noindent\textbf{Static-time Data Alignment.}
Static-time data allocation has a long history of research.  
For example, HPF (High Performance Fortran) provides compiler directives to specify data alignment among processors~\cite{hpf20}. 
Although our mechanism shares the same philosophy with the HPF directives such as {\tt block} or {\tt cyclic}, they are different in the sense that the HPF directives are applied at virtual address space, whereas it is done in the physical memory space in our mechanism since the source of non-uniformity of memory access pattern is caused when a virtual page is mapped to the physical memory domain. 

%\vspace{0.40em}
\noindent\textbf{Multiple GPUs.}
Static-time data allocation has also been researched in the context of multiple GPUs. 
A system with multiple GPUs is more close to an MPI-based system, since each GPU has its own memory and physical address space is not interleaved across multiple GPU memories. 
In this sense, several algorithms were proposed to automatically partition data among multiple devices, e.g., multiple GPUs or CPUs and GPUs~\cite{cab:vil14, lee:sam13, kim:kim11, luk:hon09, lee:sam15, gre:obo11, ram:bon13}.
In contrast, the focus of work work is to enable data partitioning among memory stacks via selective use of coarse-grain interleaving (hardware mechanism) and to enable co-location of computations with the data they access (software mechanism). 
%\vspace{-0.5em}
\section{Conclusion}
We introduce {\mech} that realizes co-location of computation and data in a system with multiple near-data processing (NDP) memory stacks.
We make an observation that the key for efficient use of NDP in improving performance and energy efficiency is to reduce remote data accesses.
To this end, we first propose a lightweight hardware mechanism that supports dual-mode address mapping at a page granularity.
With this, a page can either be spread across memory stacks (for data shared by SMs in more than one memory stacks or data primarily used by the host processor) or localized to a single memory stack (for the data exclusively used by SMs in one memory stack).
Second, we propose a software/hardware cooperative mechanism that (1) identifies exclusively accessed pages based on the anticipated access pattern for each data structure, and (2) steers computations to the memory where data they access is located. 
To anticipate the access pattern for each memory object, we utilize a combination of compile-time analysis and profiler-assisted techniques.
We use the LLVM infrastructure and perform the symbolic analysis for pattern detection. 
To co-locate computations and data, we propose an affinity-based work scheduling algorithm.
Our extensive evaluations across a wide range of workloads show that {\mech} improves performance by 31\% and reduces 38\% remote data accesses over a baseline system that cannot exploit compute-data affinity characteristics.

\clearpage

\bibliographystyle{IEEEtranS}
\bibliography{ref}

% Generated by IEEEtranS.bst, version: 1.13 (2008/09/30)
\begin{thebibliography}{10}
\providecommand{\url}[1]{#1}
\csname url@samestyle\endcsname
\providecommand{\newblock}{\relax}
\providecommand{\bibinfo}[2]{#2}
\providecommand{\BIBentrySTDinterwordspacing}{\spaceskip=0pt\relax}
\providecommand{\BIBentryALTinterwordstretchfactor}{4}
\providecommand{\BIBentryALTinterwordspacing}{\spaceskip=\fontdimen2\font plus
\BIBentryALTinterwordstretchfactor\fontdimen3\font minus
  \fontdimen4\font\relax}
\providecommand{\BIBforeignlanguage}[2]{{%
\expandafter\ifx\csname l@#1\endcsname\relax
\typeout{** WARNING: IEEEtranS.bst: No hyphenation pattern has been}%
\typeout{** loaded for the language `#1'. Using the pattern for}%
\typeout{** the default language instead.}%
\else
\language=\csname l@#1\endcsname
\fi
#2}}
\providecommand{\BIBdecl}{\relax}
\BIBdecl

\bibitem{hpf20}
``{High Performance Fortran Language Specification},'' 1997.

\bibitem{intel_sw}
``{Intel® 64 and IA-32 Architectures Software Developer’s Manual},'' 2011.

\bibitem{hbm2}
\emph{High Bandwidth Memory (HBM)}, 2015.

\bibitem{ahn:hon15}
J.~Ahn, S.~Hong, S.~Yoo, O.~Mutlu, and K.~Choi, ``{A scalable
  processing-in-memory accelerator for parallel graph processing},'' in
  \emph{Proceedings of the 42nd Annual International Symposium on Computer
  Architecture (ISCA)}, 2015.

\bibitem{ahn:yoo15}
J.~Ahn, S.~Yoo, O.~Mutlu, and K.~Choi, ``{PIM-enabled Instructions: A
  Low-overhead, Locality-aware Processing-in-memory Architecture},'' in
  \emph{Proceedings of the 42nd Annual International Symposium on Computer
  Architecture (ISCA)}, 2015.

\bibitem{aki:fra15}
B.~Akin, F.~Franchetti, and J.~C. Hoe, ``{Data reorganization in memory using
  3D-stacked DRAM},'' in \emph{Proceedings of the 42nd Annual International
  Symposium on Computer Architecture (ISCA)}, 2015.

\bibitem{bat:sum16}
T.~Barrett, S.~Mediratta, T.~jun Kwon, R.~Singh, S.~Ch, J.~Sondeen, and
  J.~Draper, ``{A Double-Data Rate (DDR) Processing-in-Memory (PIM) Device with
  WideWord Floating-Point Capability},'' in \emph{2006 IEEE International
  Symposium on Circuits and Systems (ISCAS)}, 2006.

\bibitem{cab:vil14}
J.~Cabezas, L.~Vilanova, I.~Gelado, T.~B. Jablin, N.~Navarro, and W.-m. Hwu,
  ``{Automatic Execution of single-GPU Computations Across Multiple GPUs},'' in
  \emph{2014 23rd International Conference on Parallel Architecture and
  Compilation Techniques (PACT)}, 2014.

\bibitem{cha:dev94}
R.~Chandra, S.~Devine, B.~Verghese, A.~Gupta, and M.~Rosenblum, ``{Scheduling
  and Page Migration for Multiprocessor Compute Servers},'' in
  \emph{Proceedings of the Sixth International Conference on Architectural
  Support for Programming Languages and Operating Systems (ASPLOS)}, 1994.

\bibitem{che:boy09}
S.~Che, M.~Boyer, J.~Meng, D.~Tarjan, J.~W. Sheaffer, S.-H. Lee, and
  K.~Skadron, ``{Rodinia: A Benchmark Suite for Heterogeneous Computing},'' in
  \emph{Proceedings of the 2009 IEEE International Symposium on Workload
  Characterization (IISWC)}, 2009.

\bibitem{chu:jay13}
M.~Chu, N.~Jayasena, D.~P. Zhang, and M.~Ignatowski, ``{High-level Programming
  Model Abstractions for Processing in Memory},'' in \emph{WoNDP: 1st Workshop
  on Near-Data Processing}, 2013.

\bibitem{far:ahn15}
A.~Farmahini-Farahani, J.~H. Ahn, K.~Morrow, and N.~S. Kim, ``{NDA: Near-DRAM
  acceleration architecture leveraging commodity DRAM devices and standard
  memory modules},'' in \emph{2015 IEEE 21st International Symposium on High
  Performance Computer Architecture (HPCA)}, 2015.

\bibitem{gha:jal16}
M.~Ghasempour, A.~Jaleel, J.~D. Garside, and M.~Luj{\'{a}}n, ``{DReAM: Dynamic
  Re-arrangement of Address Mapping to Improve the Performance of DRAMs},'' in
  \emph{The International Symposium on Memory Systems (MEMSYS)}, 2016.

\bibitem{gok:hol95}
M.~Gokhale, B.~Holmes, and K.~Iobst, ``{Processing in Memory: The Terasys
  Massively Parallel PIM Array},'' \emph{Computer}, 1995.

\bibitem{gre:obo11}
D.~Grewe and M.~F.~P. O'Boyle, ``{A Static Task Partitioning Approach for
  Heterogeneous Systems Using OpenCL},'' in \emph{Proceedings of the 20th
  International Conference on Compiler Construction: Part of the Joint European
  Conferences on Theory and Practice of Software (CC/ETAPS)}, 2011.

\bibitem{hal:kog99}
M.~Hall, P.~Kogge, J.~Koller, P.~Diniz, J.~Chame, J.~Draper, J.~LaCoss,
  J.~Granacki, J.~Brockman, A.~Srivastava, W.~Athas, V.~Freeh, J.~Shin, and
  J.~Park, ``{Mapping Irregular Applications to DIVA, a PIM-based
  Data-Intensive Architecture},'' in \emph{Proceedings of the 1999 ACM/IEEE
  Conference on Supercomputing (SC)}, 1999.

\bibitem{har:fer09}
N.~Hardavellas, M.~Ferdman, B.~Falsafi, and A.~Ailamaki, ``{Reactive NUCA:
  Near-optimal Block Placement and Replication in Distributed Caches},'' in
  \emph{Proceedings of the 36th Annual International Symposium on Computer
  Architecture (ISCA)}, 2009.

\bibitem{hsi:ebr16}
K.~Hsieh, E.~Ebrahimi, G.~Kim, N.~Chatterjee, M.~O’Connor, N.~Vijaykumar,
  O.~Mutlu, and S.~W. Keckler, ``{Transparent Offoading and Mapping (TOM):
  Enabling Programmer-Transparent Near-Data Processing in GPU Systems},'' in
  \emph{Proceedings of the 43nd Annual International Symposium on Computer
  Architecture (ISCA)}, 2016.

\bibitem{kan:hua99}
Y.~Kang, W.~Huang, S.-M. Yoo, D.~Keen, Z.~Ge, V.~Lam, P.~Pattnaik, and
  J.~Torrellas, ``{FlexRAM: toward an advanced intelligent memory system},'' in
  \emph{2012 IEEE 30th International Conference on Computer Design (ICCD)},
  1999.

\bibitem{kim:kim13}
G.~Kim, J.~Kim, J.~H. Ahn, and J.~Kim, ``{Memory-centric system interconnect
  design with Hybrid Memory Cubes},'' in \emph{Proceedings of the 22Nd
  International Conference on Parallel Architectures and Compilation Techniques
  (PACT)}, 2013.

\bibitem{kim:kim11}
J.~Kim, H.~Kim, J.~H. Lee, and J.~Lee, ``{Achieving a single compute device
  image in OpenCL for multiple GPUs},'' in \emph{Proceedings of the 16th ACM
  Symposium on Principles and Practice of Parallel Programming (PPoPP)}, 2011.

\bibitem{lee:sam13}
J.~Lee, M.~Samadi, Y.~Park, and S.~Mahlke, ``{Transparent CPU-GPU Collaboration
  for Data-parallel Kernels on Heterogeneous Systems},'' in \emph{Proceedings
  of the 22Nd International Conference on Parallel Architectures and
  Compilation Techniques (PACT)}, 2013.

\bibitem{lee:sam15}
J.~Lee, M.~Samadi, Y.~Park, and S.~Mahlke, ``{SKMD: Single Kernel on Multiple
  Devices for Transparent CPU-GPU Collaboration},'' \emph{ACM Transactions on
  Computer Systems (TOCS)}, 2015.

\bibitem{luk:hon09}
C.-K. Luk, S.~Hong, and H.~Kim, ``{Qilin: exploiting parallelism on
  heterogeneous multiprocessors with adaptive mapping},'' in \emph{Proceedings
  of the 42nd Annual IEEE/ACM International Symposium on Microarchitecture
  (MICRO)}, 2009.

\bibitem{mur:kog00}
R.~C. Murphy, P.~M. Kogge, and A.~Rodrigues, ``{The Characterization of Data
  Intensive Memory Workloads on Distributed PIM Systems},'' in \emph{Revised
  Papers from the Second International Workshop on Intelligent Memory Systems
  (IMS)}, 2000.

\bibitem{nai:had17}
L.~Nai, R.~Hadidi, J.~Sim, H.~Kim, P.~Kumar, and H.~Kim, ``{GraphPIM: Enabling
  Instruction-Level PIM Offloading in Graph Computing Frameworks},'' in
  \emph{2017 IEEE 23rd International Symposium on High Performance Computer
  Architecture (HPCA)}, 2017.

\bibitem{lif:yin15}
L.~Nai, Y.~Xia, I.~G. Tanase, K.~Hyesoon, and C.-Y. Lin, ``{GraphBIG:
  Understanding Graph Computing in the Context of Industrial Solutions},'' in
  \emph{International Conference for High Performance Computing, Networking,
  Storage and Analysis (SC)}, 2015.

\bibitem{nai:ant15}
R.~Nair, S.~F. Antao, C.~Bertolli, P.~Bose, J.~R. Brunheroto, T.~Chen, C.~Y.
  Cher, C.~H.~A. Costa, J.~Doi, C.~Evangelinos, B.~M. Fleischer, T.~W. Fox,
  D.~S. Gallo, L.~Grinberg, J.~A. Gunnels, A.~C. Jacob, P.~Jacob, H.~M.
  Jacobson, T.~Karkhanis, C.~Kim, J.~H. Moreno, J.~K. O'Brien, M.~Ohmacht,
  Y.~Park, D.~A. Prener, B.~S. Rosenburg, K.~D. Ryu, O.~Sallenave, M.~J.
  Serrano, P.~D.~M. Siegl, K.~Sugavanam, and Z.~Sura, ``{Active Memory Cube: A
  processing-in-memory architecture for exascale systems},'' \emph{IBM Journal
  of Research and Development}, 2015.

\bibitem{osk:cho98}
M.~Oskin, F.~T. Chong, and T.~Sherwood, ``{Active Pages: a computation model
  for intelligent memory},'' in \emph{Proceedings of the 25th Annual
  International Symposium on Computer Architecture (ISCA)}, 1998.

\bibitem{pat:and97-2}
D.~Patterson, T.~Anderson, N.~Cardwell, R.~Fromm, K.~Keeton, C.~Kozyrakis,
  R.~Thomas, and K.~Yelick, ``{A case for intelligent RAM},'' \emph{IEEE
  Micro}, 1997.

\bibitem{pat:and97}
D.~Patterson, T.~Anderson, N.~Cardwell, R.~Fromm, K.~Keeton, C.~Kozyrakis,
  R.~Thomas, and K.~Yelick, ``{Intelligent RAM (IRAM): chips that remember and
  compute},'' in \emph{IEEE International Solids-State Circuits Conference
  (ISSCC)}, 1997.

\bibitem{pet:dan16}
P.~Pessl, D.~Gruss, C.~Maurice, M.~Schwarz, and S.~Mangard, ``{DRAMA:
  Exploiting DRAM Addressing for Cross-CPU Attacks},'' in \emph{25th USENIX
  Security Symposium (USENIX Security 16)}, 2016.

\bibitem{pug:jes14}
S.~H. Pugsley, J.~Jestes, H.~Zhang, R.~Balasubramonian, V.~Srinivasan,
  A.~Buyuktosunoglu, A.~Davis, and F.~Li, ``{NDC: Analyzing the impact of
  3D-stacked memory+logic devices on MapReduce workloads},'' in \emph{IEEE
  International Symposium on Performance Analysis of Systems and Software
  (ISPASS)}, 2014.

\bibitem{ram:bon13}
T.~Ramashekar and U.~Bondhugula, ``{Automatic Data Allocation and Buffer
  Management for multi-GPU Machines},'' \emph{ACM Transactions on Architecture
  and Code Optimization (TACO)}, 2013.

\bibitem{rod:hem11}
A.~F. Rodrigues, K.~S. Hemmert, B.~W. Barrett, C.~Kersey, R.~Oldfield,
  M.~Weston, R.~Risen, J.~Cook, P.~Rosenfeld, E.~CooperBalls, and B.~Jacob,
  ``{The Structural Simulation Toolkit},'' \emph{ACM SIGMETRICS Performance
  Evaluation Review}, 2011.

\bibitem{ros:cop11}
P.~Rosenfeld, E.~Cooper-Balis, and B.~Jacob, ``{DRAMSim2: A Cycle Accurate
  Memory System Simulator},'' \emph{IEEE Computer Architecture Letters (CAL)},
  2011.

\bibitem{parboil}
J.~A. Stratton, C.~Rodrigues, I.-J. Sung, N.~Obeid, L.-W. Chang, N.~Anssari,
  G.~D. Liu, and W.~mei W.~Hwu, ``{Parboil: A Revised Benchmark Suite for
  Scientific and Commercial Throughput Computing},'' \emph{IMPACT Technical
  Report}, 2012.

\bibitem{zha:jay13}
D.~P. Zhang, N.~Jayasena, A.~Lyashevsky, J.~Greathouse, M.~Meswani, M.~Nutter,
  and M.~Ignatowski, ``{A New Perspective on Processing-in-memory Architecture
  Design},'' in \emph{Proceedings of the ACM SIGPLAN Workshop on Memory Systems
  Performance and Correctness (MSPC)}, 2013.

\bibitem{zha:jay14}
D.~Zhang, N.~Jayasena, A.~Lyashevsky, J.~L. Greathouse, L.~Xu, and
  M.~Ignatowski, ``{TOP-PIM: Throughput-oriented Programmable Processing in
  Memory},'' in \emph{Proceedings of the 23rd International Symposium on
  High-performance Parallel and Distributed Computing (HPDC)}, 2014.

\bibitem{zha:zhu00}
Z.~Zhang, Z.~Zhu, and X.~Zhang, ``{A Permutation-based Page Interleaving Scheme
  to Reduce Row-buffer Conflicts and Exploit Data Locality},'' in
  \emph{Proceedings of the 33rd Annual ACM/IEEE International Symposium on
  Microarchitecture (MICRO)}, 2000.

\bibitem{jur:mur95}
J.~H. Zurawski, J.~E. Murray, and P.~J. Lemmon, ``{The Design and Verification
  of the AlphaStation 600 5-series Workstation},'' \emph{Digital Technical
  Journal}, 1995.

\end{thebibliography}
\end{document}